\documentclass{jfm_modified_2021}
\usepackage{graphicx}
\usepackage{amsmath, amssymb}
\usepackage[inactive,tightpage]{preview}
\usepackage{microtype}
\usepackage[normalem]{ulem}
\usepackage{mathtools,bm}
\usepackage{esvect}
\usepackage{enumitem}
\usepackage{import}

\pdfmapfile{+rsfso.map}
\DeclareSymbolFont{rsfso}{U}{rsfso}{m}{n}
\DeclareSymbolFontAlphabet{\mathscr}{rsfso}

\usepackage{pdflscape}
\usepackage{afterpage}
\usepackage{flafter}
\usepackage{rotating}
\usepackage{fancyhdr}
\fancypagestyle{lscape}{%
\fancyhf{} 
\fancyfoot[LE]{}
\fancyfoot[LO] {}
 
}

\usepackage{bm}
\usepackage{microtype}

\usepackage{hyperref}
\usepackage[usenames, dvipsnames]{xcolor}
\hypersetup{colorlinks=true,linkcolor=MidnightBlue,citecolor=MidnightBlue}
\usepackage{natbib}
\usepackage{array}
\usepackage{booktabs}
\usepackage{multirow}
\usepackage{cleveref}

\usepackage{caption}
\usepackage[position=top]{subcaption}
\usepackage{tabularx}
\usepackage{siunitx}
\newcolumntype{Y}{>{\centering\arraybackslash}X}

\usepackage{tikz}
\usepackage{pgfplots}
\usetikzlibrary{backgrounds, calc, patterns.meta, shadows, arrows}
\usepackage[outline]{contour} 
\contourlength{0.15em}
\usetikzlibrary{external}

\usepackage{grffile}
\pgfplotsset{compat=newest}
\usetikzlibrary{plotmarks}
\usetikzlibrary{arrows.meta}
\usepgfplotslibrary{patchplots}

\usepackage[outline]{contour} 
\contourlength{0.15em}

\usepackage{lscape}

\newcommand*{\ep}{\epsilon}

\newcommand*{\Oh}{\mathcal{O}}

\newcommand*{\de}{\operatorname{d\!}{}} 
\newcommand{\dd}[2]{\frac{\de#1}{\de#2}}

\usepackage{printlen}

\shorttitle{Asymptotic analysis of catchment models}
\shortauthor{Morawiecki \& Trinh}

\title{On the development and analysis of coupled surface-subsurface models of catchments.
\\[0.3em] \large Part 2. A three-dimensional benchmark model and its properties}

\author{Piotr Morawiecki\corresp{\email{piotr.morawiecki@bath.edu}}
 \and Philippe H. Trinh\corresp{\email{p.trinh@bath.ac.uk}}}

\affiliation{
    Department of Mathematical Sciences, University of Bath, Bath BA2 7AY, UK
}

\pubyear{XXXX}
\volume{YYY}
\pagerange{xxx--xxx}
\date{\today~[Draft]}

\newcommand{\editt}[1]{#1}
\newcommand{\edit}[1]{#1}

\def\d{\mathrm{d}}
\newcommand{\dx}[1]{\frac{\partial #1}{\partial x}}

\newcommand{\dy}[1]{\frac{\partial #1}{\partial y}}

\newcommand{\dz}[1]{\frac{\partial #1}{\partial z}}

\newcommand{\dt}[1]{\frac{\partial #1}{\partial t}}

\newcommand{\qin}[0]{q_\mathrm{in}}
\newcommand{\dxhat}[1]{\frac{\partial #1}{\partial \hat{x}}}
\newcommand{\dyhat}[1]{\frac{\partial #1}{\partial \hat{y}}}
\newcommand{\dzhat}[1]{\frac{\partial #1}{\partial \hat{z}}}
\newcommand{\betazx}[0]{\beta_{zx}}
\newcommand{\betazy}[0]{\beta_{zy}}

\newcommand*{\Nop}{\mathcal{N}}

\makeatletter
\newcommand\Label[1]{&\refstepcounter{equation}(\theequation)\ltx@label{#1}&}
\makeatother

\begin{document}

\maketitle


\begin{abstract}
    
    \noindent The objective of this three-part work is to formulate and rigorously analyse a number of reduced mathematical models that are nevertheless capable of describing the hydrology at the scale of a river basin (\emph{i.e.} catchment). Coupled surface and subsurface flows are considered. 
    
    In this second part, we construct a benchmark catchment scenario and investigate the effects of parameters within their typical ranges. Previous research on coupled surface-subsurface models have focused on numerical simulations of site-specific catchments. Here, our focus is broad, emphasising the study of general solutions to the mathematical models, and their dependencies on dimensionless parameters. This study provides a foundation based on the examination of a geometrically simple three-dimensional benchmark scenario. We develop a nondimensional coupled surface-subsurface model and extract the key dimensionless parameters.
    \edit{Asymptotic methods demonstrate under what conditions the model can be reduced to a two-dimensional form, where the principal groundwater and overland flows occur in the hillslope direction. Numerical solutions provide guidance on the validity of such reductions, and demonstrate the parametric dependencies corresponding to a strong rainfall event.}
\end{abstract}



\section{Introduction}
\label{sec:introduction}

\noindent Since the publication of the Stanford Watershed Model by \cite{crawford1966digital} a wide range of computational models of catchment-scale hydrology have been  developed \citep{singh2003watershed}. Indeed, over two hundred models have been identified in the extensive review by \cite{peel2020historical}. 

Such computational models are primarily designed in order to predict the evolution of surface and subsurface flow in a particular river basin given the input precipitation via rainfall or snowfall. These so-called \textit{rainfall-runoff models} are often divided into three classes: empirical, conceptual, and physical \citep{sitterson2018overview}; this last category of physical models involves those that are developed from the known physical principles of hydrodynamics. For instance, the Richards equation is commonly used to model the subsurface flow through the saturated or unsaturated soil, while the Saint Venant equation is used to model the overland and the channel flow. For a detailed introduction, see \cite{shaw2015hydrology}. Such governing equations form the foundation of many currently used computational integrated catchment models, \emph{e.g.} MIKE SHE \citep{abbott1986introduction1, abbott1986introduction2}, HydroGeoSphere \citep{brunner2012hydrogeosphere}, ParFlow \citep{kollet2006integrated}, and OpenGeoSys \citep{kolditz2012opengeosys}.

\edit{However, in contrast to computational studies, there seems to have been more limited work on the systematic mathematical analysis of the fundamental principles of coupled surface-subsurface catchment-scale models. A proper mathematical formulation can allow us to better understand the importance of parameters, establish the limits of simplifications used in computational models, and develop analytical or semi-analytical solutions in certain scenarios.}

\subsection{On the development and benchmarking of computational models}

\noindent The Stanford Watershed Model IV is a conceptual model, which is considered to be amongst the earliest attempts to computationally model the entire hydrological cycle. Its publication resulted in the subsequent development of an enormous number of independent computational models \citep{donigian2006history}. However, further computational power was needed before the first physically-based models were implemented. Notable early examples include TOPMODEL \citep{kirkby1979physically}, MIKE SHE \citep{abbott1986introduction2, abbott1986introduction1}, and IHDM (Institute of Hydrology Distributed Model, cf. \citealt{beven1987institute}).

The abundance of independent catchment models results in a need to better understand their accuracy and differences. Within the industry, such models are typically assessed by comparing model's predictions (usually after earlier calibration) to available data, such as river flow or groundwater depth measurements [see a detailed introduction to rainfall-runoff modelling by \cite{beven2011rainfall}]. However, there is criticism, \emph{e.g.} by \cite{hutton2016most}, that the models in hydrology are often not reproducible. \citet[p.~6]{beven2018hypothesis} highlighted some fundamental issues that continue to exist in the state-of-the-art of catchment modelling. He noticed that:
\begin{quotation}
    \noindent \textit{``Where model intercomparisons have been done, different models give different results, and it is often the case that the rankings of models in terms of performance will vary with the period of data used, site, or type of application. This would seem to be a very unsatisfactory situation for the advancement of the science, especially when we expect that when true predictions are made, they will turn out to be at best highly uncertain and at worst quite wrong."}
\end{quotation}

\noindent In response to this problem, many numerical methodologies for calibration, cross-validation, and uncertainty estimation have been developed (see {\itshape e.g.} \citealt{beven1992future} and \citealt{gupta2006model}). These methods allow us to assess, in a more unbiased way, the accuracy of the models. However, they do not necessarily point out the reason for potential inaccuracies. As \cite{kirchner2006getting} argued, advancing the science of hydrology requires developing not only models that match the available data, but models that are theoretically justified.

\begin{figure}
    \centering
    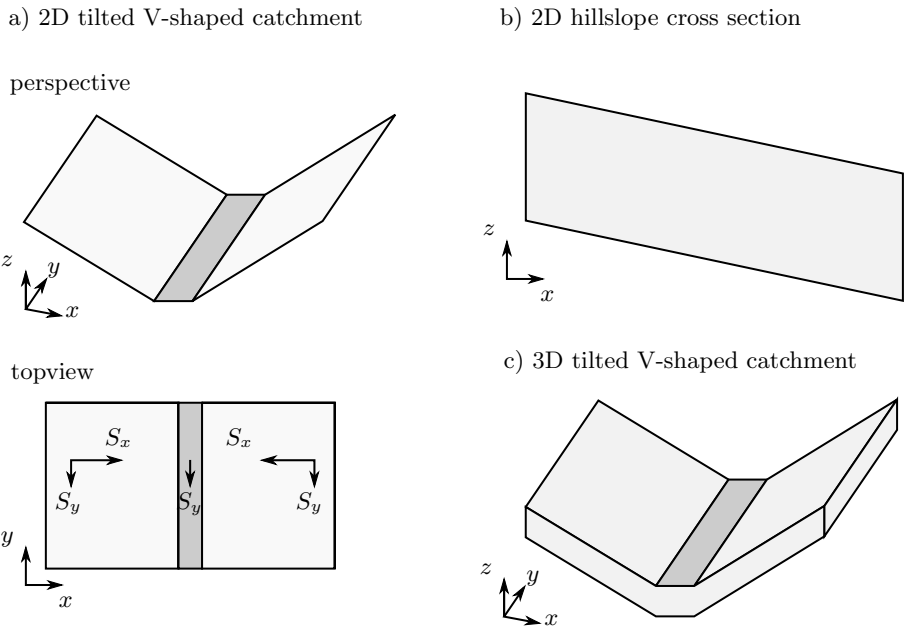
    \caption{Illustration of the idealised catchment geometries developed in the works of \cite{kollet2006integrated} and \cite{gilbert2016global}. Geometries (a) and (c) represent a tilted V-shaped river valley with two hillslopes and a river in the middle, with the latter geometry introducing subsurface flow in the third dimension. Geometry (b) represents a single two-dimensional hillslope with a river channel located at the right boundary.}
    \label{fig:benchmark_geometries}
\end{figure}

Independently, there has been an effort to develop simple (idealised) catchment geometries that can be used as benchmarks to assess the accuracy of integrated catchment models in fully controlled conditions. \cite{kollet2006integrated} used a tilted V-shaped catchment geometry (\cref{fig:benchmark_geometries}a) to compare predictions for overland flow given by four different hydrological catchment models with an analytical one-dimensional solution. Then, they introduced a simple two-dimensional hillslope (\cref{fig:benchmark_geometries}b), which they used to explore the sensitivity of an integrated ParFlow model for geometry settings (\emph{e.g.} water table depth, hydraulic conductivity, and soil heterogeneities). The same benchmark scenarios were used by \cite{sulis2010comparison} to compare ParFlow and CATHY models \citep{bixio2000physically}. This study was followed by far more extensive intercomparison studies by \cite{maxwell2014surface} and \cite{kollet2017integrated}, which used these and other benchmark scenarios to compare the results obtained using a wide range of integrated catchment models.

In the meanwhile, simple catchment/hillslope scenarios have also been used to assess coupled surface and subsurface flow with other models---this includes examination of evapotranspiration \citep{kollet2009influence}, atmosphere \citep{sulis2017coupling}, biochemistry \citep{cui2014modeling}, the impact of climate change \citep{markovich2016hydrogeological}, and the effects of different types of heterogeneities, \emph{e.g.} the heterogeneity of the land surface \citep{rihani2015isolating}, soil properties \citep{meyerhoff2011quantifying}, and even flow through fractures \citep{sweetenham2017assessing}. The two studies by \cite{jefferson2015active} and \cite{gilbert2016global} introduced a three-dimensional tilted V-shaped catchment with a constant soil depth (\cref{fig:benchmark_geometries}c). The authors used this geometry to perform a sensitivity analysis of integrated catchment models---the first study by \cite{jefferson2015active} focused on the energy flux terms, while the second by \cite{gilbert2016global} studied the heterogeneity of soil permeability. In both studies, the sensitivity analysis results were used to obtain a certain level of dimensionality reduction by applying the active subspace method \citep{constantine2015active}.

An open question remains, though, whether one can simplify the model and its parameter space based on the analysis of the governing equations (even in a simplified catchment scenario), rather than based on the numerical results; this could provide more rigorous insight into the limits of applicability of the above computational reductions.


\edit{Another aspect we shall investigate in this work concerns the study of key nondimensional parameters characterising surface-subsurface hydrological processes. We highlight some prior works that have used nondimensionalisation in order to analyse governing equations describing \emph{individual} flow components:} for example, this has been applied by \cite{akan1985similarity} in the Saint Venant equations to study the water infiltration into the ground. It has also been used by \emph{e.g.} \cite{warrick1990analytical, warrick1993scaling, haverkamp1998scaling} for the study of the one-dimensional Richards equation, describing water vertical infiltration through the unsaturated soil.

A notable work, in which nondimensionalisation plays a prominent role for the case of coupled surface-subsurface models, was performed by \cite{sivapalan1987hydrologic}, and focuses on the TOPMODEL scheme of \cite{kirkby1979physically}. A similar study was performed by \cite{calver1991dimensionless} for the IHDM model \citep{beven1987institute}. In particular, \cite{calver1991dimensionless} define a list of ten dimensionless parameters, study the dependencies between selected parameters, and discuss the properties of the hydrographs. However, the relevant scale of dimensionless parameters is not assessed in this latter work.

\subsection{On the development of a simple benchmark model}

\noindent The modern-day catchment hydrology is studied based on the simulation of complex integrated catchment models. So far, however, the authors have not found many comprehensive studies on the design and analysis of simple benchmark scenarios for coupled surface-subsurface catchment models. Our work in Part 1 \citep{paper1} has initiated this task via a thorough examination of the typical parameter sizes. In this part, we focus on the design of a three-dimensional benchmark, study its typical dynamics, and discuss its reduction to lower-dimensional models. 

Compared to the existing literature, there are three novel elements in our study:

\begin{enumerate}[label={(\roman*)},leftmargin=*, align = left, labelsep=\parindent, topsep=3pt, itemsep=2pt,itemindent=0pt ]
\item Our benchmark scenario is posed on a simple geometry, but the surface/subsurface governing equations are posed in a general three-dimensional dimensionless form.

\item We use the dimensionless model to provide a rigorous argument behind the simplifications commonly used in computational hydrology. We discuss the reduction of a problem geometry to 2D in detail, and comment on the kinematic/dynamic wave approximation. We achieve this by setting clear conditions on the size of dimensionless parameters, and justify them based on the typical values of model parameters obtained in the previous part of our work (see table 1 from Part 1).

\item We use the benchmark model to numerically explore the impact of the remaining parameters on the system in response to intensive rainfall. Because we attempt to do this in a systematic and analytical way, this work also serves to set a more rigorous benchmark standard for future studies. For example, scaling laws are derived that may serve as a benchmark for other model schemes.
\end{enumerate}

\noindent \edit{Note that our study is restricted to modelling the formation of storm flow during an intensive rainfall \citep{guerin2019stream}; however, similar benchmark scenarios can be considered in order to study other flow regimes. This may include, for instance, drought flow observed during a period without any rainfall \citep{brutsaert1977regionalized}, or a sudden drawdown drainage when a rapid change of water level occurs at the outlet \citep{sanford1993hillslope}.}

We start by formulating a three-dimensional benchmark scenario in \cref{sec:Model_formulation}, which is non-dimensionalised in \cref{sec:governing_equations}. In \cref{sec:reduced_models} we show that this model can be reduced to a two-dimensional form \edit{by neglecting the subsurface and overland flow component in the y-direction}. Following the numerical methodology from \cref{sec:numerical_methods}, this model simplification is numerically assessed in \cref{sec:3D_to_2D_verification}. The impact of each parameter in the resulting two-dimensional model is summarised in \cref{sec:parameters_impact}, which is followed by the discussion in \cref{sec:discussion}.

\textbf{Symbols.} There are many symbols in this work. For ease of reference, we provide a list of symbols in \cref{tab:list1} and \cref{tab:list2} in \cref{sec:listofsymbols}.

\section{Formulation of a simplified three-dimensional catchment model}
\label{sec:Model_formulation}

\noindent In this section, we formulate a simplified catchment model, inspired by the infiltration-excess, saturation-excess, and tilted V-shaped catchment scenarios from the benchmark study by \cite{maxwell2014surface}.

\begin{figure}
    \centering
    \def\svgwidth{0.8\linewidth}
    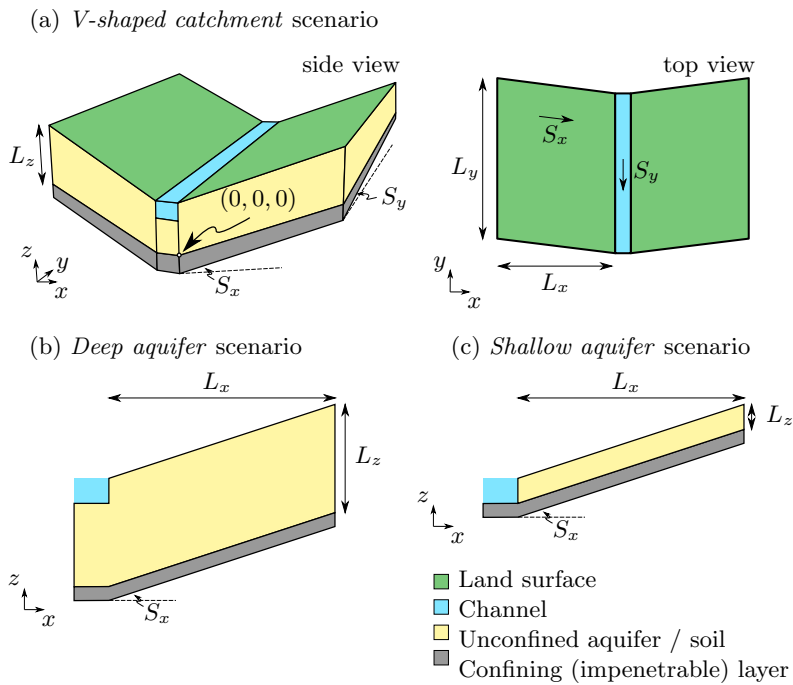
    \caption{Simplified catchment geometry in the considered scenarios (not to scale).} 
    \label{fig:simplified_catchment}
\end{figure}

\edit{We introduce the following three scenarios, as depicted in \cref{fig:simplified_catchment}.
\begin{enumerate}[label={(\roman*)},leftmargin=*, align = left, labelsep=\parindent, topsep=3pt, itemsep=2pt,itemindent=0pt ]
    \item[(a) \emph{The V-shaped catchment.}] This scenario, shown in \cref{fig:simplified_catchment}(a), represents a V-shaped catchment with a thick aquifer, where subsurface water is transferred both through the soil and through the underlying bedrock. \editt{The aquifer dimensions are $L_x\times L_y\times L_z$, where $L_z$ is the thickness of the permeable layer of the aquifer. The elevation gradient along the hillslope is denoted as $S_x$, and along the direction of the river as $S_y$.} \edit{Similar to the V-shaped scenario studied by \cite{maxwell2014surface}, we shall assume that the channel has a constant width, $w$, and zero depth, $d=0$.} Later in \cref{sec:reduced_models}, we demonstrate that under certain conditions, the scenario reduces to largely two-dimensional dynamics along the hillslope.
    \item[(b) \emph{The deep aquifer.}] This scenario, shown in \cref{fig:simplified_catchment}(b), represents a two-dimensional hillslope with a thick aquifer, where the subsurface water is transferred through both the soil and the underlying bedrock. Following the infiltration- and saturation-excess scenarios discussed in \cite{maxwell2014surface}, the channel is assumed to have a rectangular $xz$ cross-section with width, $w$, and depth,  $d$.
    \item[(c) \emph{The shallow aquifer.}] This scenario, shown in \cref{fig:simplified_catchment}(c), represents a catchment with a low-productive aquifer, in which the subsurface water is transferred only through a thin soil layer. Mathematically, the geometry of the problem is equivalent to the \textit{deep aquifer} scenario with $L_z\ll L_x$. We analyse this scenario in Part 3.
\end{enumerate}}

The focus of work in this Part 2 is the study of the \textit{V-shaped catchment} scenario and its reduction to a two-dimensional \textit{deep aquifer} scenario. In Part 3, we shall demonstrate that under the additional restrictions of the \textit{shallow aquifer} scenario, further analysis can be performed through a long wavelength reduction. In the \textit{V-shaped catchment} scenario, an orthogonal coordinate system $(x, y, z)$ is chosen such that $z$ is vertical and $y$ is directed along the channel. Using the reflection symmetry of the catchment, we can describe the catchment behaviour by only considering a hillslopes only on one side of the river.

When formulating the governing equations for overland and subsurface flow, we are going to use a more convenient non-orthogonal coordinate system, where the axes $(\hat{x}, \hat{y}, \hat{z})$ are directed along the hillslope edges. Hence, $\hat{x}$ is directed along the hillslope ($\hat{x}=0$ representing the location of the channel), $\hat{y}$ along the channel ($\hat{y}=0$ representing the location of the outlet), and $\hat{z}$ vertically ($\hat{z}=0$ representing the bottom of the aquifer). After the coordinate transformation, the entire catchment can be represented as a cuboid of dimensions $L_{\hat x} \times L_y \times L_z$. The following coordinate transformation is used:
\begin{equation}
    \label{eq:tilted_coordinates}
        x = \hat{x} \sqrt{1-\left(\frac{S_y}{S_x}\right)^2}, \qquad 
        y = \hat{x} \frac{S_y}{S_x} + \hat{y}, \qquad
        z = S_x \hat{x} + S_y \hat{y} + \hat{z}.
\end{equation}
We introduced $L_{\hat x}$ to represent the catchment width along the $\hat x$ direction given as
\begin{equation}
    L_{\hat x}=\frac{L_x}{\sqrt{1-\left(\frac{S_y}{S_x}\right)^2}}.
\end{equation}

\edit{The land surface in this geometry corresponds to
\begin{equation}
    H_\mathrm{surf}(\hat x,\hat y)=z\left(\hat{x},\hat{y},\hat{z}=L_z\right)=S_x \hat{x} + S_y \hat{y} + L_z.
\end{equation}
}

Note that real-world systems are characterised by different levels of heterogeneity of the surface, soil, and parent material properties. Here, in order to construct a minimal model, we consider properties to be homogeneous; this is similar to the assumptions made by \cite{maxwell2014surface}. Thus, the surface is assumed to have uniform roughness, and the properties of soil and rock layer are assumed to be homogeneous, \emph{i.e.} have a uniform hydraulic conductivity and water-retention curve. Also, we assume that the soil and bedrock do not include the presence of macropores and fractures, which would lead to the formation of preferential flow---see more in the reviews by \cite{bouma1981soil} and \cite{neuzil1981flow}. Because of the last assumption, the model may not properly represent the infiltration through the unsaturated zone in many of the real-world systems. As noted \emph{e.g.} by \cite{beven2013macropores}, including these effects in the model may significantly affect the timescale of infiltration.

\subsection{Asymptotic limits of geometrical parameters}%
\label{sub:asymptotic_limits_of_geometrical}

\noindent It is convenient to discuss the asymptotic limits of the key non-dimensional parameters that characterise the geometry. First, we have the slope ratios between the channel and hillslope directions, 
\begin{equation}
	\ep = \frac{S_y}{S_x},
\end{equation}
which for a typical UK catchment is \edit{$\epsilon\in [0.13,0.25]$}\footnote{\edit{The estimates represent the interquartile range based on parameters characterising over $1200$ UK catchments, values of which were estimated in Part 1 (here the first quartile is $0.13$, and the third quartile is $0.25$).}}. We also have the aspect ratio between the catchment height and the catchment dimension along the river, 
\begin{equation}
	\betazy = \frac{L_z}{L_y},
\end{equation}
which for a typical UK catchment is \edit{$\betazy\in [0.0007, 0.025]$}. Finally, we have the aspect ratio between the catchment height and the catchment length along the hillslope:
\begin{equation}
	\betazx = \frac{L_z}{L_{\hat{x}}},
\end{equation}
which for a typical UK catchment is \edit{$\betazx\in [0.1,2.1]$}. Note that as $\betazy/\betazx \to 0$, we get long catchments with a width much shorter than their length, while for $\betazy/\betazx \to \infty$, we get short catchments with a width much longer than their length.

\begin{figure}
    \centering
    \includegraphics[trim=7 0 0 0,clip]{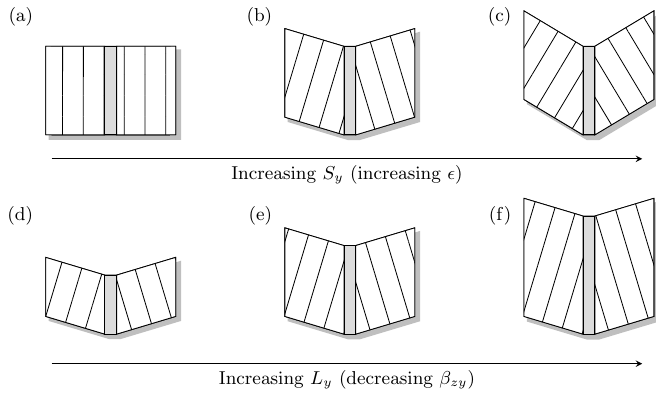}
    \caption{These illustrations provide a guide to understand the impact of changing values of the slope, $S_y$ (a--c), and length, $L_y$ (d--f) on our V-shaped catchment geometry (the channel is shaded). Lines of constant elevation of the topography are represented with dashed lines, drawn on top of a projection of the catchment onto $z = 0$. By the definition of a catchment, the top and bottom boundaries are perpendicular to lines of constant elevation (since an unperturbed flow will follow lines of the steepest descent). These dashed lines help to visualise the geometry of the later contour plots.
    \label{fig:catchment_outline}}
\end{figure}

The impact of these two parameters on the catchment geometry is schematically presented in \cref{fig:catchment_outline}. Here, we draw lines of constant topographic elevation on a projection of the catchment onto $z = 0$. Note that, for example in \cref{fig:catchment_outline}(a) for $S_y = 0$, surface and subsurface flow will typically occur in the $x$ direction, perpendicular to the river direction. In contrast, for \cref{fig:catchment_outline}(c), we may expect to observe a significant flow component parallel to the river direction.

It is important to remember that, since our interest is in the study of the benchmark model, we are not necessarily limited to studying only physical regimes. That is, it is still interesting to study the asymptotic limits so that we can establish the qualitative trends.

\subsection{Relationship to Maxwell et al. (2004)}
\label{sec:relationship_to_Maxwell}

\noindent Here, we briefly outline how the scenarios introduced above relate to the scenarios presented in the benchmark analysis of \cite{maxwell2014surface}.

In sections 4.1 and 4.2, \cite{maxwell2014surface} introduce two scenarios called the infiltration- and saturation-excess, respectively. \edit{In the infiltration scenario, precipitation exceeds the saturated soil conductivity ($r>K_s$). Only part of the precipitation infiltrates through the soil, while the remaining part accumulates at the surface to form an overland flow (the so-called Horton overland flow). In the saturation-excess scenario ($r<K_s$), overland flow is not generated unless the entire soil becomes fully saturated.}

Both scenarios are posed on a single hillslope, which represents a thin layer of soil ($L_z=5$ m) with a slope following the $x$ direction, while the river is assumed to have a fixed surface water height. \edit{Thus, this geometry represents the \textit{shallow aquifer} scenario shown in \cref{fig:simplified_catchment}, where the flow takes place only in a thin layer of the soil. Note that this geometry does not include water infiltration to the deeper permeable layers of the parent material (as in the \textit{deep aquifer} scenario in \cref{fig:simplified_catchment}), which is an effect that characterises the majority of the real-world aquifers (note a small area of aquifers without the groundwater on the UK map in fig. 4 from Part 1).}

A second limitation of the geometries considered by Maxwell is that there is no slope along the river, which drives the flow down the river valley. Although the authors included the slope perpendicular to the hillslope in a separate scenario introduced in their section 4.3 (V-shaped catchment), this benchmark scenario does not include subsurface modelling; therefore, the water infiltration into the soil was not studied. 

Our scenarios in this work combine the above two elements, \emph{i.e.} groundwater flow through deep aquifers and slope both perpendicular and along the river. Therefore, we consider a V-shaped catchment with an additional $z$-dimension allowing the saturation to vary with depth, as in the hillslope scenario. The need to introduce a tilted coordinate system comes from the fact that the elevation gradient (determining the direction of surface flow) is not perpendicular to the river, since it must have a small component along the $y$-axis. In order to satisfy the no-surface flow boundary condition at the catchment boundary, the bottom and top boundaries of the hillslope are thus inclined by a small angle,  $\phi=\mathrm{asin}(S_y/S_x))$, relative to the rectangular domain in the infiltration and saturation-excess scenarios.

Last but not least, we use the typical catchment parameters as estimated in Part 1; note that these values can be significantly different from those numerical values used in the work of \cite{maxwell2014surface}. \edit{Based on our simulations, we observed that if one were to use the parameter values given by \cite{maxwell2014surface}, this would lead to unrealistic steady states, where the seepage covers almost the entire catchment (even for relatively low levels of mean precipitation).}

\section{Governing equations (dimensional)}
\label{sec:governing_equations}

\noindent We begin with the dimensional model. \edit{As introduced in sec. 2 of Part 1,} we consider three types of flow: the subsurface flow (the 3D Richards equation), the overland flow (the 2D Saint Venant equations), and the channel flow (1D Saint Venant equation). In this section, we present governing equations for each of the flow components \edit{in our benchmark scenario}, together with the corresponding boundary conditions. General reviews of these governing equations can be found in the works of \cite{farthing2017numerical, schaake1975surface} and references therein.

\subsection{3D Richards equation for the subsurface flow}%
\label{sub:subsurface_flow}

\noindent The subsurface flow $\mathbf{q_g}(x,y,z,t)$ depends on the pressure head $h_g(x,y,z,t)$. Its evolution in time $t$ is commonly modelled using a three-dimensional Richards equation (see \emph{e.g.} \cite{dogan2005saturated} and \cite{weill2009generalized}), which is given by:
\begin{equation}
    \label{eq:Richards}
    \frac{\d\theta}{\d h_g}\dt{h_g}=\nabla\cdot \mathbf{q_g}, \quad \text{where} \quad \mathbf{q_g} = K_s K_r(h_g)\nabla(h_g+z).
\end{equation}
Here, $\nabla=(\dx{},\dy{},\dz{})$ is a standard nabla operator, $K_s > 0$ is the saturated soil conductivity and $\frac{\d\theta(h_g)}{\d h_g}$ is the so-called specific moisture capacity. We assume that the volumetric water content $\theta(h_g)$, and relative hydraulic conductivity $K_r(h_g)$ are functions of the pressure head given by the Mualem-van Genuchten (MvG) model \citep{van1980closed}:
\begin{subequations}
    \begin{align}
    \theta(h_g) &=
    \begin{cases} 
        \theta_r+\frac{\theta_s-\theta_r}{\left(1+\left(\alpha_\mathrm{MvG}h_g\right)^n\right)^m} & h_g<0 \\
        \theta_s & h_g\ge0
    \end{cases},
    \label{eq:GMtheta} \\
    K_r(h_g) &=
    \begin{cases}
        \frac{\left(1-\left(\alpha_\mathrm{MvG}h_g\right)^{n-1}\left(1+\left(\alpha_\mathrm{MvG}h_g\right)^n\right)^{-m}\right)^2}{\left(1+\left(\alpha_\mathrm{MvG}h_g\right)^n\right)^{m/2}} & h_g<0 \\
        1 & h_g\ge0
    \end{cases}.
    \label{eq:GMk}
    \end{align}
\end{subequations}
\edit{Here, the value of $h_g=0$ corresponds to the pressure head at the groundwater table surface, which separates the fully-saturated zone ($h_g>0$) from the partially-saturated zone ($h_g<0$) (see the later \cref{fig:2D_BCs}a for a reference image).} In essence, the MvG model describes the key hydraulic properties of the soil, hydraulic conductivity and saturation as nonlinear functions of the pressure head $h_g$. The model introduces further parameters $\alpha_\mathrm{MvG}$, $\theta_r$, $\theta_s$, $n$, and $m=1-\frac{1}{n}$, which depend on the soil properties. The residual water content $\theta_r$ and saturated water content $\theta_s$ represent the lowest and the highest water content, respectively. The $\alpha_\mathrm{MvG}$ parameter in $\left[\mathrm{m}^{-1}\right]$ represents the scaling factor for the pressure head $h_g$ [m]. The $n$ coefficient describes the pore sizes distribution.

\subsection{2D Saint Venant equations for the overland flow}%
\label{sub:surface_flow}

\noindent If the precipitation exceeds the inflow into the soil, water can accumulate on the surface and form overland flow. Typically, and following \emph{e.g.} \cite{tayfur1994spatially} and \cite{liu2004two} this flow is described using the two-dimensional Saint Venant equations that govern the overland water height, $h_s(x, y, t)$. \edit{Following the discussion in sec. 2.2 of Part 1, we consider mass and momentum conservation. Firstly, the continuity equation} is given by
\begin{equation}
    \label{eq:St_Venant_overland}
    \dt{h_s}=\nabla\cdot\mathbf{q_s}(h_s)+R_\mathrm{eff}-I,
\end{equation}
\noindent where $I = I(x,y,t)$ is the infiltration rate, and $R_\text{eff} = R(x,y,t) - ET(x,y,t)$ is the effective precipitation rate, which we define as the difference between the precipitation rate, $R$, and the evapotranspiration rate,  $ET$. 

The flux, $\mathbf{q_s}$, that appears in the Saint Venant equation \eqref{eq:St_Venant_overland}, is commonly obtained in hydrology using an empirical relationship known as Manning's law. Written in vector form, it is given by
\begin{equation}
    \label{eq:2D_Manning}
    \mathbf{q_s} = \frac{1}{n_s}h_s^{5/3}\frac{\mathbf{S}_f}{\sqrt{\left|\mathbf{S}_f\right|}},
\end{equation}
\noindent where $n_s$ is an empirically determined value known as Manning's coefficient, and describes the overland surface roughness; $\mathbf{S_f}$ is a dimensionless friction slope defined as gradient of energy of water per unit weight.

When Manning's law in \eqref{eq:2D_Manning} is substituted into the continuity equation \eqref{eq:St_Venant_overland}, this yields a single equation for the two unknowns, $h_s$ and $\mathbf{S}_f$. \edit{In general, the friction slope, $\mathbf{S_f}$, is given by momentum conservation [cf. eqn (2.7) in Part 1]. However, in computational integrated catchment models, a kinematic approximation is often used which neglects all effects on $\mathbf{S}_f$ other than gravity. This approximation is used in \emph{e.g.} Parflow \citep{maxwell2009parflow}, though there are others such as \emph{e.g.} MIKE SHE that implement a more complete, diffusive approximation \citep{MIKE2017manual}. In the case of the kinematic approximation,
\begin{equation}
    \label{eq:kinematic_approximation}
    \mathbf{S}_f\sim\mathbf{S}_0
\end{equation}
where $\mathbf{S}_0=-\nabla H_\mathrm{surf}$ is the elevation gradient. In this paper, we shall adopt the above kinematic approximation. This reduction significantly simplifies the problem since, under this approximation, the overland flows only down the hillslope (the $\hat{x}$-direction). As \cite{vieira1983conditions} argues, this approximation may give inaccurate predictions when the system is close to reaching a steady state.}

\subsection{1D Saint Venant equation for the channel flow}%
\label{sub:channel_flow}

\noindent Finally, we need to formulate the governing equation for the surface flow in a rectangular channel of width $w$. The channel is directed along the $\hat{y}$-axis. Following \cite{vieira1983conditions} and \cite{chaudhry2007open}, the channel flow is modelled as a one-dimensional Saint Venant equation that governs the channel water height, $z = h_c(\hat{y}, t)$, and is given by:
\begin{equation}
    \label{eq:St_Venant_channel}
    w\dt{h_c} = \qin - \dyhat{q_c},
\end{equation}
where $w(h_c, \hat{x})$ is the channel width (constant in the case of a rectangular channel), and $q_\mathrm{in}$ is a source term governing the total surface and subsurface inflow into the river. As for the overland equations, the flux, $q_c$, is assumed to be given by the empirical Manning's law, which takes the form:
\begin{equation}
    \label{eq:Manning_flow_general}
    q_c = A\frac{\sqrt{S_f^\mathrm{river}}}{n_c}\left(\frac{A}{P}\right)^{2/3},
\end{equation}
where $A$ is the channel cross-section, $P$ is the channel wetted perimeter, $n_c$ is Manning's coefficient dependent on banks and channel bed roughness, and $S_f^\mathrm{river}$ is the friction slope, \edit{which under kinematic approximation is equal to the elevation gradient along the river:
\begin{equation}
    \label{eq:friction_slope_channel}
    S_f^\mathrm{river}=S_y.
\end{equation}}

\edit{In summary, the solution of the channel flow involves the substitution of Manning's equation \eqref{eq:Manning_flow_general} and the friction slope \eqref{eq:friction_slope_channel} into the Saint Venant equation \eqref{eq:St_Venant_channel}. For the case of the V-shaped catchment illustrated in \cref{fig:simplified_catchment}, where there is a rectangular channel, this involves setting the area $A = wh_c$ and $P = w + 2 h_c$.}

\edit{The above channel flow model, when coupled to the hillslope forms a challenging numerical computation due to the nonlinearity. Instead, for the purpose of numerical computation, we apply a model simplification of the channel flow similar to what is considered by \cite{maxwell2014surface}. In this simplification, we set $A=wh_c$ and approximate $P \approx w$, and hence ignore the friction effects of the channel side walls. In this case, Manning's equation \eqref{eq:Manning_flow_general} becomes
\begin{equation} \label{eq:Manning_flow}
    q_c = w\frac{\sqrt{S_y}}{n_c}h_c^{5/3}.
\end{equation}
The later simulations will thus involve the solution of the Saint Venant equation \eqref{eq:St_Venant_channel} with the shallow Manning's equation \eqref{eq:Manning_flow} and \eqref{eq:friction_slope_channel}. The advantage of the above approximation is that the channel flow problem satisfies a similar partial differential equation to the surface-flow, but with adjusted coefficient values.
}

\subsection{Boundary conditions}
\label{sec:BC}

\begin{figure}
    \centering
    \def\svgwidth{\linewidth}
    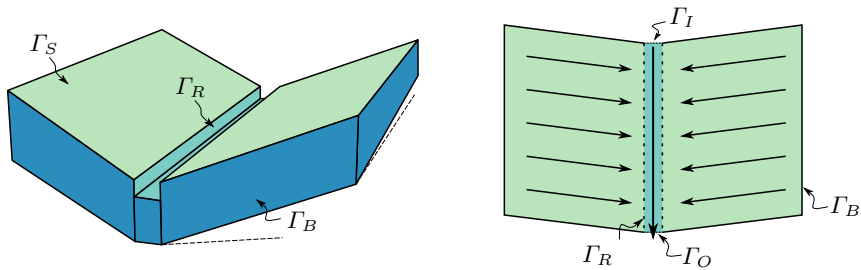
    \caption{Boundaries defined for the V-shaped tilted catchment.}
    \label{fig:boundaries}
\end{figure}

\noindent The domain consists of four types of boundaries: (i) the catchment boundary, $\Gamma_B$, both for the surface and subsurface part of the domain, including the bedrock constraining the aquifer from the bottom; (ii) the land surface $\Gamma_s$; (iii) the river bank, $\Gamma_R$; (iv) the river outlet, $\Gamma_O$; and (v) the river inlet $\Gamma_I$ (see \cref{fig:boundaries}).

\begin{enumerate}[label={(\roman*)},leftmargin=*, align = left, labelsep=\parindent, topsep=3pt, itemsep=2pt,itemindent=0pt ]
\begin{subequations}
\item \edit{Firstly, there is no surface flow through the catchment boundary. Also, for simplicity, we will assume that there is no groundwater flow through this boundary -- the rainwater can only leave the catchment via the channel flow.} Hence, on $\Gamma_B$, we set no-flow conditions for both subsurface and surface flow:
\begin{equation}
    \label{eq:bo_flow_BC}
    \mathbf{q_g}\cdot\mathbf{n} = 0, \quad \mathbf{q_s}\cdot\mathbf{n} = 0\qquad \text{on $\Gamma_B$}.
\end{equation}
\edit{Alternatively, one could introduce a free-flow condition for the groundwater flow, $\mathbf{q_g}\cdot\mathbf{n} = 0$, to allow for the outflow of the groundwater flow through the catchment boundary. In this work, we have chosen no-flow conditions to guarantee that the entirety of the rainfall eventually reaches the channel, which simplifies the resultant water balance.}
\item Next, on the land surface, $\Gamma_s$, continuity of pressure and flow between the groundwater and surface water yields:
\begin{equation}
\label{eq:surface_BC}
        h_s =
        \begin{cases}
            0 & \quad \text{if } h_g<0\\
            h_g  & \quad \text{if } h_g>0
        \end{cases} \qquad \text{and} \qquad
        \mathbf{q_g}\cdot\mathbf{n} = I \qquad \text{on $\Gamma_s$}.
\end{equation}
This first condition imposes continuity of pressure only if the groundwater reaches $\Gamma_s$, while the second imposes the condition of rain infiltration, $I$. 
\item \edit{On the river bank, $\Gamma_R$, we also impose continuity of pressure between the channel water, which is characterised by a hydrostatic profile, $h(z)=h_c-z$, and the subsurface pressure head, $h_g$:
\begin{equation}
    \label{eq:BC_channel}
    h_g = h_c - z \qquad \text{on $\Gamma_R$}.
\end{equation}}
\item \edit{At the inlet, located at the upstream end of the river, $\Gamma_I$, we can impose an inflow from the upstream part of the catchment, which is located outside of the modelled domain. In general, it can change over time, and so
\begin{equation}
    \label{eq:BC_inlet_flow}
    q_c = q_\mathrm{input}(t), \qquad \text{on $\Gamma_I$}.
\end{equation}
In our benchmark scenario, we assume for simplicity that $q_\mathrm{input}=0$, as if the top boundary represents the start of the stream. Such a stream is referred to as a \textit{first-order stream} (see \citealt{strahler1957quantitative}), \editt{however in the real-world situations the first order stream does not reach the catchment divide. The presented model can be also generalised to represent higher-order streams by including a non-zero upstream inlows $q_\mathrm{input}(t)$.}}
\end{subequations}
\end{enumerate}

\edit{Note that the kinematic approximation \eqref{eq:kinematic_approximation} that we follow in our work reduces the overland and channel equations to advective equations, rather than advective-diffusion equations. Thus in this approximation, the downstream boundary conditions -- at the river bank $\Gamma_R$ (for overland flow) and at the catchment outlet $\Gamma_O$ (for channel flow) -- do not have to be imposed.}

This means that effectively the channel flow does not impact the overland flow. However, overland flow impacts the channel flow thought the inflow term $\qin$ in \eqref{eq:St_Venant_channel}. According to flow continuity, the input to the channel flow is the sum of the overland flow and the total groundwater flow, integrated over the entire channel perimeter at the given cross-section. Thus
\begin{equation}
    \label{eq:river_inflow}
    \qin = \mathbf{q_s}\big|_{\Gamma_R}\cdot\mathbf{n} + \int_{\Gamma_R} \mathbf{q_g}\cdot\mathbf{n}\;\d l.
\end{equation}

\edit{Two-way coupling between channel flow and subsurface flow is maintained via boundary condition \eqref{eq:BC_channel}, and two-way coupling between the overland flow and subsurface flow is maintained via \eqref{eq:surface_BC}.}

\subsection{Initial conditions of the benchmark}

\noindent The choice of the initial condition is more arbitrary. In contrast to the benchmark scenarios by \cite{maxwell2014surface}, which assumed a constant groundwater depth, we select a more realistic setting, where the groundwater profile is given by its typical shape for a given catchment. Thus, we find a steady state of $h_g(x,y,z)$, $h_s(x,y)$, and $h_c(x,y)$ given by the time-independent versions of the governing equations~\eqref{eq:Richards}, \eqref{eq:St_Venant_overland}, and \eqref{eq:St_Venant_channel}, solved for a given mean precipitation rate~$R_\mathrm{eff}=R_0$:
\begin{equation}
    \label{eq:IC}
    \nabla\cdot \mathbf{q_g} = 0,  \qquad \nabla\cdot \mathbf{q_s} + R_0 - I = 0, \quad\text{and}\quad \edit{\qin - \dyhat{q_c}=0}.
\end{equation}
Once this initial state is found by solving the above system of equations, we then explore the evolution of $h_g(x,y,z,t)$ and $h_s(x,y,t)$ caused by intensive rainfall, $R_\mathrm{eff}>R_0$, which moves the system away from the initial state.

\section{Governing equations (nondimensional)}

\subsection{Nondimensionalisation}
\label{sec:nondimensionalisation}

\noindent
The governing equations for subsurface, surface, and channel flow presented in \cref{sec:Model_formulation} are now written in tilted coordinates $(\hat x, \hat y, \hat z)$ [cf. \eqref{eq:tilted_coordinates}] and given in dimensional form in \cref{app:governing_dim}. In order to understand the relative size of the terms appearing in the governing equations, we nondimensionalise these equations. The following scalings are used:
\begin{equation*}
\begin{aligned}[c]
        \hat{x} &= L_{\hat x} \hat{x}', \\
        \hat{y} &= L_y \hat{y}', \\
        \hat{z} &= L_z \hat{z}',
\end{aligned}
\quad
\begin{aligned}[c]
        h_g &= L_z h_g', \\
        h_s &= L_s h_s', \\
        h_c &= L_c h_c',
\end{aligned}
\quad
\begin{aligned}[c]
        t &= t_0 t', \\
        \theta(h) &= \theta'(h'), \\
        K_r(h) &= K_r'(h').
\end{aligned}
\quad
\begin{aligned}[c]
        R_\mathrm{eff} &= r R_\mathrm{eff}', \\ 
        I &= r I', \\
        \qin &= r L_{\hat x} \qin',
\end{aligned}
\end{equation*}
Here, $r$ is an average value of $R_\mathrm{eff}$. We shall choose the characteristic time, $t_0$, overland water height, $L_s$, and channel water height, $L_c$, according to:
\begin{equation} \label{eq:nondimpack1}
        t_0 = \frac{L_z}{K_s}, \qquad
        L_s = \left(\frac{L_{\hat x} n_s r}{S_x^{1/2}}\right)^{3/5}, \qquad
        L_c = \left(\frac{n_c r L_{\hat x} L_y}{w S_y^{1/2}}\right)^{3/5}.
\end{equation}
The choice of the above quantities comes from balancing the leading terms in the governing equations for subsurface, overland, and channel flow respectively. Their formulation, in terms of tilted coordinates, is presented in eqns \eqref{eq:Richards_xyz_form}, \eqref{eq:2D_saint_Venant_xy_form}, and \eqref{eq:St_Venant_rectangular_channel} in \cref{app:governing_dim}.


Additionally, the non-dimensional terms in \eqref{eq:nondimpack1} have straightforward physical interpretations. The timescale, $t_0$, describes a characteristic time that rainwater needs to penetrate the aquifer of thickness $L_z$, infiltrating with a characteristic speed $K_s$ (such flow occurs due to gravity if there is no hydraulic gradient, \emph{e.g.} during uniform rainfall). The quantity $L_s$ represents the height of the overland flow at the river bank in a steady state with rainfall $r$ (assuming that the entire rainfall forms an overland flow, \emph{i.e.} no infiltration appears). Similarly, $L_c$ is an approximate height of the flow in a wide channel at the river outlet in a steady state. Crucially, we note that the choice of the above scaling seems to be correct for our chosen benchmark, with all relevant dimensionless quantities of typical order unity in the numerical simulations of \cref{sec:3D_time_dependent_solution}.

\edit{It should be noted that even though $t_0$ is a characteristic time of the vertical flow through the soil, other timescales are present. For example, we shall observe typically shorter timescales for the overland flow, and much longer timescales for the horizontal flow through the soil. Further discussion of the separation of timescales appears in Part 3 of our work.} 

\subsection{Summary of governing equations and parameters}

\noindent We collect the nondimensional governing equations from \cref{app:governing_dimless}. To review, our hydrological problems in the three-dimensional geometry consist of solving three time-dependent partial differential equations for three unknowns: (i) a 3D Richards equation for the subsurface flow~\eqref{eq:dimless_subsurface_main}; (ii) a 2D Saint Venant equation for the overland flow~\eqref{eq:dimless_overland_main}; and (iii) a 1D Saint Venant equation for the channel flow~\eqref{eq:dimless_channel_main}. In the tilted frame, these are respectively
\edit{\begin{subequations} \label{eq:dimless_altogether}
    \begin{align}
        \label{eq:dimless_subsurface_main}
        \text{(Subsurface)} && \frac{\d\theta}{\d h}\bigg\rvert_{h=h_g}\dt{h_g} &=
        \Nop_1(h_g) + \betazy^2 \Nop_2(h_g) + \epsilon \betazy \Nop_3(h_g), \\
        \label{eq:dimless_overland_main}
        \text{(Overland)} && \tau_s\dt{h_s} &= \dxhat{}\left(h_s^{5/3}\right) + R_\mathrm{eff} - I, \\
        \label{eq:dimless_channel_main}
        \text{(Channel)} && \tau_c\dt{h_c} &= \qin - \dyhat{}\left(h_c^{5/3}\right).
    \end{align}
\end{subequations}
where the subsurface equations involve operators definitions:
\begin{subequations}
    \begin{align}
    \Nop_1(h_g) =&
    \begin{aligned}[t]\dzhat{} \left[K_r(h_g) \left(\dzhat{h_g}+1\right)\right]
    + \betazx S_x\dxhat{} \left[K_r(h_g) \left(2\dzhat{h_g}+1\right)\right] \\
    + \betazx^2 \left(1 + S_x^2\right)\dxhat{} \left[K_r(h_g) \dxhat{h_g}\right] - \frac{\d\theta}{\d h}\bigg\rvert_{h=h_g}\dt{h_g},
    \end{aligned} \\
    \Nop_2(h_g) =&  \left(1 + S_y^2\right) \dyhat{} \left[K_r(h_g)\dyhat{h_g}\right], \\
    \Nop_3(h_g) =& 2 \betazx \left(1+S_x^2\right) \dxhat{} \left[K_r(h_g)\dyhat{h_g}\right] + S_x\dyhat{} \left[K_r(h_g) \left(2\dzhat{h_g}+1\right)\right].
    \end{align}
\end{subequations}
Expressions for $\theta(h_g)$ and $K_r(h_g)$ are provided in \cref{app:governing_dimless}.} Each partial differential equation in \eqref{eq:dimless_altogether} is solved subject to boundary conditions posed on the domain boundaries given by \eqref{eq:bc_nondim_first}--\eqref{eq:bc_nondim_last}.

Finally, these equations are characterized by nine independent dimensionless parameters, \{$\betazx$, $\betazy$, $\sigma_x$, $\sigma_y$, $\tau_s$, $\tau_c$, $\gamma$, $\alpha$, $\rho$\}, with definitions provided in \cref{app:list_of_parameters}.

\section{Model reduction to a two-dimensional model}
\label{sec:reduced_models}

\noindent In \cref{sec:Model_formulation}, we formulated a general three-dimensional catchment model. The purpose of this section is to discuss the nondimensionalisation of the model, which subsequently allows for the determination of the key dimensionless parameters governing the system. Once these are known, we may use the typical dimensional values established in Part 1 in order to compare the relative strengths of the various physical effects of the system.

We highlight two approximations:
\begin{enumerate}[label={(\roman*)},leftmargin=*, align = left, labelsep=\parindent, topsep=3pt, itemsep=2pt,itemindent=0pt ]
    \item Considering either \editt{small river slope ($S_y\ll S_x$), short ($L_y\ll L_z$), or long catchment ($L_y\gg L_z$) approximations together with the necessary low channel limit ($L_c\ll L_z$)}, we may reduce the general three-dimensional governing equations for $h_g$ and $h_s$ to a two-dimensional form neglecting the flow along the $y$-axis.
    \item In addition, \editt{in the case of the \textit{shallow aquifer} scenario ($L_z\ll L_x$)}, we may apply a shallow-water approximation to further reduce the 2D hillslope model to a 1D model.
\end{enumerate}
\noindent In this section, the approximations given in (i) are discussed. The regime of (ii) and its consequences are explored in Part 3 of our work. 

\edit{\subsection{Discussion of the low channel height limit, \texorpdfstring{$L_c\ll L_z$}{}}}
\label{sec:low_channel_height}

\noindent \edit{The three-dimensional model in $(x,y,z)$ can be formally approximated by a two-dimensional model in $(x,z)$ if the subsurface profile, $h_g(x,y,z,t) \sim h_{g,0}(x,z,t)$, and the surface profile, $h_s(x,y,t) \sim h_{s,0}(x,t)$, and both asymptotic approximations are consistent with the initial and boundary conditions (at the leading order).}

\edit{We observe that the boundary condition \eqref{eq:BC_channel}, along the channel, $\Gamma_R$, depends in general on the channel water height, $h_c(y,t)$, which can vary along the catchment. For example, the dimensional height can vary from $h_c=0$ at $y=L_y$ (if there is no inflow to the river from the upstream point) to a dimensional height $h_c=L_c$ at $y = 0$ (in the case of the steady-state outflow). Returning to nondimensional values for $h_g$, $h_c$,, and $z$, \eqref{eq:BC_channel} yields
\begin{equation}
    \label{eq:BC_channel_dimless}
    h_g = \frac{L_c}{L_z}h_c - z, \qquad \text{on $\Gamma_R$}.
\end{equation}
Typically, the values of $L_c/L_z$ are very small: based on the UK catchment data from Part 1, we can extract the interquartile range for $L_c/L_z$, namely $[0.0011, 0.0147]$ (\emph{i.e.} the middle half of UK catchments have $L_c/L_z$ within this interval). Thus, even though the channel water height may vary along the channel, it is negligibly small comparing to the typical variation of the pressure head. In the limit of $L_c/L_z \to 0$, we see that the subsurface boundary condition is
\begin{equation}
    h_g \sim -z, \qquad \text{on $\Gamma_R$},
\end{equation}
which is no longer $y$-dependent.}

\subsection{An asymptotic expansion for small river slopes, in  \texorpdfstring{$\epsilon = S_y/S_x$}{}}
\label{sec:asymptotic_expansion_for_small_eps}

\noindent \edit{Although the remaining boundary conditions [\eqref{eq:bo_flow_BC}--\eqref{eq:BC_inlet_flow} without \eqref{eq:BC_channel}] are not explicitly $y$-dependent, the solution $h(x,z,t)$ may still exhibit leading $y$-dependent effects due to \emph{e.g.} the topography. However, there are certain approximations in which these effects are very small---for example, when the slope along the channel $S_y$ is much lower than the slope along the hillslope $S_x$.} \edit{Note that the aspect ratio introduced in \cref{sub:asymptotic_limits_of_geometrical}, $\epsilon = S_y/S_x$, typically has small values (half of UK catchments have $\epsilon$ between $0.13$ and $0.25$). Here we shall demonstrate that when $\epsilon \ll 1$ (equivalent to $S_y\ll S_x$), the solution is expected to be predominantly two-dimensional.}

Firstly, we rewrite the set of dimensionless governing equations for the subsurface and overland flows, \eqref{eq:dimless_subsurface} and~\eqref{eq:dimless_overland}, in a simpler form highlighting its structure:
\begin{subequations} \label{eq:governing_eqs_structure}
\begin{align}
    \text{(Subsurface)} && \frac{\d\theta}{\d h}\bigg\rvert_{h=h_g}\dt{h_g} &=
    \Nop_1(h_g) + \betazy^2 \Nop_2(h_g) + \epsilon \betazy \Nop_3(h_g), \\
    \text{(Overland)} && \tau_s\dt{h_s} &= \dxhat{}\left(h_s^{5/3}\right) + R_\mathrm{eff} - I,
    \label{eq:overland_asymptotic_sec}
\end{align}
\end{subequations}
where the nonlinear operators, $\Nop_i$, for $i = 1, 2, 3$ are defined in~\eqref{eq:Nop_definitions} 
in \cref{app:governing_dimless}. Note that these operators are dependent on $h_g$ and $h_s$, and independent of $\epsilon$ and $\betazy$, which are the only dimensionless parameters involving $S_y$ and $L_y$. 

When $\epsilon = 0$, we can verify that the solutions are independent of $\hat{y}$, \emph{i.e.} they can be written as $h_g(\hat x,\hat y,\hat z)=h_{g,0}(\hat x,\hat z)$ and $h_s(\hat x,\hat y)=h_{s,0}(\hat x)$. This is caused by the combination of three facts:
\begin{enumerate}[label={(\roman*)},leftmargin=*, align = left, labelsep=\parindent, topsep=3pt, itemsep=2pt,itemindent=0pt ]
    \item term $\Nop_1(h_g)$ is independent of $\hat{y}$;
    \item operators $\Nop_2$ and $\Nop_3$ applied to a function independent of $\hat{y}$ become $0$; and
    \item the no-flux boundary condition at $\hat{y}=0, \, 1$ in  \eqref{eq:bc_nondim_first} is then:
    \begin{equation}
        \label{eq:no_flux_bc_tilted}
        \dyhat{h_g}-\frac{\epsilon}{1-\epsilon^2}\left(\frac{\betazx}{\betazy}\dxhat{h_g}-\epsilon\dyhat{h_g}-\frac{2S_x}{\betazy}\dzhat{h_g}\right) = 0.
    \end{equation}
    Hence, for $\epsilon=0$, the above boundary condition is satisfied by $h_g=h_{g,0}(\hat x,\hat z)$.
    \item \edit{From eqn \eqref{eq:overland_asymptotic_sec} only $R_\mathrm{eff}$ and $I$ can be $\hat{y}$-dependent terms, but in the considered scenario $R_\mathrm{eff}$ is constant, and $I(\hat{x},\hat{y})$ is $\hat{y}$-independent as long as $h_g$ is.}
\end{enumerate}

Essentially, $\epsilon=0$ is associated with a zero gradient along the river, \emph{i.e.} there is no forcing flow in the $\hat{y}$-direction, and the domain becomes transitionally symmetric in that direction. 

\edit{There is an important consideration in the formal limit as $S_y\propto\epsilon \to 0$. In this limit, holding other parameters fixed, the $L_c$ defined in \eqref{eq:nondimpack1} tends to $\infty$, and so the $L_c\ll L_z$ condition from
\eqref{sec:low_channel_height} is no longer satisfied. This is due to the fact that when reducing the gradient along the channel, $S_y$, the channel water height must increases in order to maintain a significant channel flow [cf. Manning's law \eqref{eq:Manning_flow}]. Therefore, we would expect for two-dimensional dynamics to dominate when 
\begin{equation} \label{eq:newlessthan}
    \left(\frac{n_c r L_{\hat x} L_y}{w L_z^{5/3}}\right)^2\ll S_y\ll S_x.
\end{equation}
Above, the expression on the left-hand side is obtained from the definition of $L_y$ from \eqref{eq:nondimpack1}, and represents the value of $S_y$, for which $L_c=L_z$. In real-world situations, $S_y$ (with a median value of $0.014$ based on the data collected in Part 1) is higher by a few orders of magnitude over this threshold (with a median value of $6.1\cdot 10^{-11}$), and so only the second approximation in \eqref{eq:newlessthan} needs to be considered.}

\subsection{Asymptotic expansions for short (\texorpdfstring{$\betazy \gg 1$}{}) and long (\texorpdfstring{$\betazy \ll 1$}{}) catchments}
\label{sec:limit_beta}

\noindent There are additional limits that allow us to reduce the three-dimensional problem into simpler two-dimensional formulations at the leading order, and these involve the non-dimensional geometrical parameter
\begin{equation}
\betazy \equiv \frac{L_z}{L_y}.
\end{equation}
For instance, in the limit as $\betazy \to \infty$, the three-dimensional catchment reduces to an infinitely thin hillslope profile with a negligible flow in the perpendicular direction to the hillslope (since we imposed no-flow conditions at $\hat{y}=0$ and $\hat{y}=1$). Equivalently, this corresponds to an asymptotically short section of a river. From \eqref{eq:no_flux_bc_tilted}, we see that the leading-order profile should satisfy the $\de{h_{g,0}}/\de{\hat{y}}=0$ condition at $\hat{y} = 0, 1$, which is automatically satisfied for a $\hat{y}$-independent solution. As argued in the previous section, we also conclude that such a $\hat{y}$-independent solution will also satisfy the governing equations~\eqref{eq:governing_eqs_structure}.

Using a similar analysis to the one presented in the previous section, by balancing leading terms in the boundary conditions~\eqref{eq:no_flux_bc_tilted}, we can show that the full three-dimensional solution can be expanded in terms of $\betazy^{-1}$:
\begin{subequations}
\begin{align}
h_g(\hat x,\hat y,\hat z) &=h_{g,0}(\hat x,\hat z)+\betazy^{-1} h_{g,1}(\hat x,\hat y,\hat z) + \Oh(\betazy^{-2}), \\
h_s(\hat x,\hat y) &= h_{s,0}(\hat x)+\betazy^{-1} h_{s,1}(\hat x,\hat y) + \Oh(\betazy^{-2}).
\end{align}
\end{subequations}

\begin{figure}
    \centering
    \includegraphics{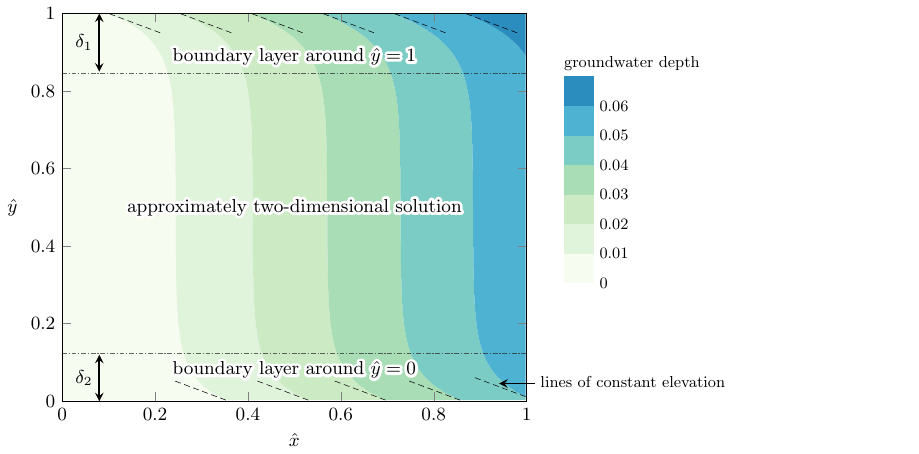}
    \caption{\edit{This graphic shows the steady-state depth of the groundwater table, according to the 3D model. We note that the solution is mostly $\hat{y}$-independent, except for two apparent boundary layers around $\hat{y}=0$ and $\hat{y}=1$; near these points, the groundwater table aligns with the lines of constant elevation. The figure is generated using the solver described in \cref{sec:numerical_methods} using the parameter values given in \cref{tab:simulation_settings}, with except for two values: we use $L_y=13680$ m ($\beta_{zy}=0.05$) and $S_y=0.0075$ ($\epsilon=0.1$); as a result, this graphic matches an inset in \cref{fig:3D_vs_2D_steady_states}. The boundary layer thicknesses, $\delta_1$ and $\delta_2$, tend to zero as $\beta_{zy} \to \infty$.}}
    \label{fig:boundary_layer_diagram}
\end{figure}

The last interesting limit we discuss is $\betazy\rightarrow 0$, which corresponds to the situation of an asymptotically long river. Similarly, the two-dimensional solution satisfies the governing equations~\eqref{eq:governing_eqs_structure}. However this time it does not satisfy the no-flow boundary condition~\eqref{eq:no_flux_bc_tilted}. Therefore, we expect to observe a boundary layer around $\hat{y}=0$ and $\hat{y}=1$ (see \cref{fig:boundary_layer_diagram}). Consequently, $h_{g,0}$ and $h_{s,0}$ are understood to represent the `outer' asymptotic solutions, valid for $0 < \hat{y} < 1$. 

Without loss of generality, let us consider the boundary layer near $\hat{y} = 0$. We rescale $\hat{y}=\delta\hat{y}'$ where $\delta(\betazy)$ is a characteristic size of the boundary layer. After applying this transformation, the governing equations \eqref{eq:governing_eqs_structure} become:
\begin{subequations}
\begin{align}
    \text{(Subsurface)} && \frac{\d\theta}{\d h}\bigg\rvert_{h=h_g}\dt{h_g} &=
    \Nop_1(h_g) + \frac{\betazy^2}{\delta^2} \Nop_2(h_g) + \epsilon \frac{\betazy}{\delta} \Nop_3(h_g), \\
    \text{(Overland)} && \tau_s\dt{h_s} &= \dxhat{}\left(h_s^{5/3}\right) + R_\mathrm{eff} - I.
\end{align}
\end{subequations}

The three-dimensional effects given by $\Nop_2$ and $\Nop_3$ become significant when the characteristic size of the boundary layer is of the order $\Oh(\betazy)$. Hence, we conclude that the thickness of the boundary layer is $\Oh(1/\betazy)$. So, if we consider the $\edit{\betazy\rightarrow 0}$ limit, the solution for the problem becomes two-dimensional except for an infinitely thin boundary layer around the boundaries.

\subsection{Summary of the two-dimensional model}

\noindent \editt{To summarise, we considered three approximations for small river slope ($\epsilon\ll 1$), short catchments ($\betazy\gg 1$), and long catchments ($\betazy\ll 1$). We showed that under any of these approximations,} the model can be approximately represented in the following two-dimensional form:
\begin{subequations}
\begin{align}
    \label{eq:2D_Richards} 
    \text{(Subsurface)} && \frac{\d\theta}{\d h}\bigg\rvert_{h=h_{g,0}}\dt{h_{g,0}} &=
    \Nop_1(h_{g,0}), \\
    \label{eq:1D_overland}
    \text{(Overland)} && \tau_s\dt{h_{s,0}} &= \dxhat{}\left(h_{s,0}^{5/3}\right)+R_\mathrm{eff}-I,
\end{align}
\end{subequations}
with
\begin{multline}
    \Nop_1(h_{g,0}) = \dzhat{} \left[K_r(h_{g,0}) \left(\dzhat{h_{g,0}}+1\right)\right]
    + \betazx S_x\dxhat{} \left[K_r(h_{g,0}) \left(2\dzhat{h_{g,0}}+1\right)\right] \\
    + \betazx^2 \left(1 + S_x^2\right)\dxhat{} \left[K_r(h_{g,0}) \dxhat{h_{g,0}}\right].
\end{multline}

\edit{Both the groundwater and overland flows reaching the channel contribute to the river flow in the $\hat{y}$-direction governed by~\eqref{eq:dimless_channel}. Thus, 
\begin{equation}   
    \label{eq:2D_channel}
    \text{(Channel)} \qquad \tau_c\dt{h_c} = \qin - \dyhat{}\left(h_c^{5/3}\right),
\end{equation}
where the inflow $\qin=\qin(t)$ is given by the total groundwater and overland inflow to the channel. This equation allows us to find the hydrograph at the outlet of the catchment $Q(t)=q_c(\hat{y}=0,t)$. However, note that after the reduction to a two-dimensional model, equations \eqref{eq:2D_Richards} and \eqref{eq:1D_overland} are uncoupled from \eqref{eq:2D_channel}, and so $\qin$ can be found without solving the last equation. Therefore, we can study the properties of the 2D model by solving equations \eqref{eq:2D_Richards} and \eqref{eq:1D_overland} alone, and exploring the properties of the river inflow term $\qin(t)$. We follow this approach in the study of the 2D model in \cref{sec:parameters_impact}, and in Part 3 of our work.}

\begin{figure}
    \centering
    \def\svgwidth{\linewidth}
    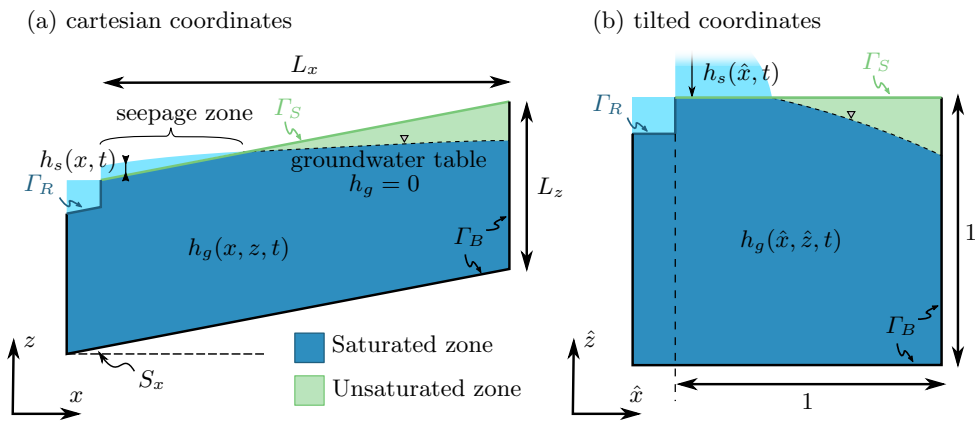
    \caption{An illustration of a two-dimensional hillslope geometry in Cartesian coordinates (left) and in the tilted coordinate system (right). \edit{These represent a cross-section (in the $xz$-plane) of the original V-shaped catchment}.}
    \label{fig:2D_BCs}
\end{figure}

The above system of PDEs forms a model describing surface and subsurface flow in a 2D hillslope cross-section, as presented in \cref{fig:2D_BCs}. The boundaries are now one-dimensional, but the boundary conditions are the same as in the three-dimensional model, as given by~\eqref{eq:bc_nondim_first}--\eqref{eq:bc_nondim_last}. As before, for the initial condition, we consider the steady state of the above system for $R_\mathrm{eff}=R_0$. In \cref{sec:3D_to_2D_verification}, we will explore the accuracy of this approximation numerically, investigating the size of three-dimensional features of the full solution and their behaviour in limits formulated above.

A two remarks are in order:

\begin{enumerate}[label={(\roman*)},leftmargin=*, align = left, labelsep=\parindent, topsep=3pt, itemsep=2pt,itemindent=0pt ]

    
    \item Firstly, there are some remaining terms in the dimensionless governing equations, which are small; however, they should not be neglected. \editt{For example, coefficient,
    \begin{equation}
        \tau_s=\frac{...}{...}=\frac{...}{...}
    \end{equation}
    is small, which means that the characteristic timescale of accumulation of surface water is much shorter than the characteristic timescale of vertical subsurface flow. However, temporal term} of~\eqref{eq:1D_overland} becomes significant for small values of $t$. Since the typical rainfall times are much lower than the  characteristic time of groundwater transfer (estimated as $t_0\approx6.8\cdot 10^7~\mathrm{s}\approx 2$ years), we are often interested in the short-time behaviour, and therefore, this term should not be neglected.
    
    \item  Secondly, the $\betazx$ term is also small in the \textit{shallow aquifer} scenario compared to the leading term representing the flow in the $\hat z$-direction, and therefore it can be neglected in regions with significant temporal effects (\emph{e.g.} in partially saturated zones impacted by the rainfall). However, in the fully saturated zone, where $h_g>0$, we have $\frac{\d\theta}{\d h}=0$. In this zone, the balance between the two remaining terms needs to be maintained---the horizontal flow becomes high enough to balance the vertical flow. Therefore, the $\betazx$ term cannot be neglected in the fully-saturated zone; however, another simplification based on the shallow-water approximation can be considered. This will be explored in the Part 3 of our work.
\end{enumerate}

\section{Numerical methodology}
\label{sec:numerical_methods}

\noindent In order to validate the reduction of the 3D model to the 2D approximation and quantify the impact of model parameters on the observed peak flow, we follow a numerical approach.

Here, we present the numerical method used to implement the coupled surface-subsurface model based on the governing equations introduced in \cref{sec:governing_equations}. \edit{To summarise, these are:
\begin{subequations} \label{eq:numerical_altogether}
    \begin{align}
        \label{eq:numerical_subsurface_main}
        \text{(Subsurface)} &&  \frac{\d\theta}{\d h_g}\dt{h_g}&=\nabla\cdot \left(K_s K_r(h_g)\nabla(h_g+z)\right), \\
        \label{eq:numerical_overland_main}
        \text{(Overland)} && 
        \dt{h_s}&=\nabla\cdot\left(\frac{1}{n_s}h_s^{5/3}\frac{\mathbf{S}_0}{\sqrt{\left|\mathbf{S}_0\right|}}\right)+R_\mathrm{eff}-I, \\
        \label{eq:numerical_channel_main}
        \text{(Channel)} &&  w\dt{h_c} &= \qin - w\frac{\sqrt{S_y}}{n_c}h_c^{5/3}.
    \end{align}
\end{subequations}
subject to boundary conditions \eqref{eq:bo_flow_BC}-\eqref{eq:BC_inlet_flow}.}

\edit{Our numerical implementation has two applications in this study. Firstly, in \cref{sec:3D_to_2D_verification} we use the numerical scheme based on a discrete version of this equations to verify our reductions to the 2D problem introduced in \cref{sec:reduced_models}. Secondly, in \cref{sec:sensitivity_analysis} we numerically analyse features of a benchmark scenario in a reduced two-dimensional analysis. We use the same equations as above, excluding $y$-dependent terms and channel flow.}
The source codes were written in MATLAB and are available in our GitHub repository \citep{github2}.

\subsection{Model discretisation}

\begin{figure}
    \includegraphics{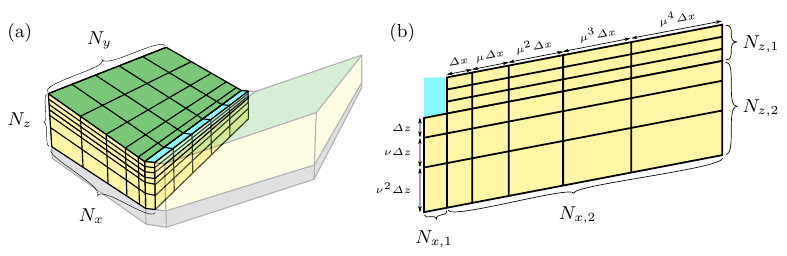}
    \caption{(a) Discretisation of the 3D catchment, representing the \textit{V-shaped catchment} scenario. In the case of the 2D \textit{deep aquifer} and \textit{shallow aquifer} model, we set $N_y=0$. (b) Example of mesh refinement. The size of edges is given by the geometric series with ratios $\mu$ and $\nu$.}
    \label{fig:hillslope_mesh}
\end{figure}

The implementation of the 3D and 2D models is based on the finite volume method. The entire hillslope is divided into $N_x\times N_y\times N_z$ cells. $N_z$ is additionally split into $N_{z,1}$ cells representing the layer of soil with the same depth as the channel, and $N_{z,2}=N_z-N_{z,1}$ cells representing deeper layers of the aquifer, as illustrated in \cref{fig:hillslope_mesh}a. In the case of the \textit{shallow aquifer} scenario, we set $N_{z,2}=0$. The implementation allows for a mesh refinement by varying the cells extent, according to a geometric series (see \cref{fig:hillslope_mesh}b).

\edit{Depending on the type of simulation, we handle the channel differently. In the case of two-dimensional simulations, we will assume the water level in the channel to be equal to the channel depth (unless stated otherwise). In the case of three-dimensional simulations, $h_c(y,t)$ will be iteratively computed. However, following the V-shaped catchment scenario by \cite{maxwell2014surface}, we consider the channel flow in the same way as the overland flow, just with a different Manning's coefficient. This way, we neglect the interactions with the river banks; however, the simulations are significantly more stable.}

A value of $h_g$ is assigned to each cell to represent $h_g$ at the centroid of the cell. Due to the pressure continuity condition at the surface, $h_s=h_g$, so there is no need to consider $h_s$ at the surface as an independent variable \edit{(the same applies to $h_c$ in three-dimensional simulations)}. One challenge is the significant difference in the timescales between the overland and groundwater flow (the ratio of which is quantified with the dimensionless parameter $\tau_s\approx2.8\cdot 10^{-4}$). A stable numerical scheme for overland flow requires a shorter time step than groundwater flow. Therefore, for each groundwater flow step, we compute several steps of the surface flow, while simultaneously satisfying the continuity boundary condition at the surface.

The groundwater in each time step is found using an implicit scheme. The following discretised version of Richards equation~\eqref{eq:dimless_subsurface} for each cell is used:
\begin{subequations} \label{eq:Richard_discretisied}
\begin{equation}
        V_i\frac{\theta\left(h_i^{\prime t+1}\right)+\frac{\d\theta}{\d h}\big\rvert_{h_i^{\prime t+1}}\left(h_i^{t+1}-h_i^{\prime t+1}\right)-\theta\left(h_i^t\right)}{\Delta t} = 
        \sum_{j\in\text{neighbours}} \mathcal{G}_{ij}, 
\end{equation}
where
\begin{multline}
\mathcal{G}_{ij} = 
 \frac{S_{i,j}}{\cos(\eta_{i,j})} 
 \left(K_{i,j}'+\frac{\d K_{i,j}'}{\d h}\left(h_{u(i,j)}^{t+1}-h_{u(i,j)}^{\prime t+1}\right)\right) \\
 \times \frac{h_j^{t+1}+z_j-h_i^{t+1}-z_i}{\|\mathbf{r}_{i\rightarrow j}\|}\frac{\boldsymbol{\beta}\cdot\mathbf{r}_{i\rightarrow j}}{\|\mathbf{r}_{i\rightarrow j}\|}.
\end{multline}
\end{subequations}

Few remarks should be made here:

\begin{enumerate}[label={(\roman*)},leftmargin=*, align = left, labelsep=\parindent, topsep=3pt, itemsep=2pt,itemindent=0pt ]

\item The left-hand side represents the estimated rate of change of the $i$\textsuperscript{th} cell's water content. Here $V_i$ is the cell's volume, $\Delta t$ is the time step duration, $h_i^t$ and $h_i^{t+1}$ is the pressure head ($h_g$) in cell $i$ at time step $t$ (previous one) and $t+1$ (current one) respectively, $h_i^{\prime t}$ is the pressure head computed in the previous iteration of the implicit scheme, and $\theta$ is the saturation given by the Mualem-Van Genuchten model~\eqref{eq:MvG_dimless}.

\item The right-hand side represents the sum of all flows between cell $i$ and its neighbours. Here, $S_{i,j}$ is the area of the face between cell $i$ and $j$, $\eta_{i,j}$ is the angle between this face and the line joining these cells' centroids, $\mathbf{r}_{i\rightarrow j}$ is the vector from the centroid of cell $i$ to the centroid of cell $j$, and $z_i$ is the $z$ coordinate of the $i$\textsuperscript{th} cell centroid. Additionally, $\boldsymbol{\beta}=(\betazx^2,\betazy^2,1)$ is a vector describing the anisotropy coming from different scaling factors in the dimensionless model. $K_{i,j}'$ is the hydraulic conductivity of the face between cell $i$ and $j$. It is computed using the upwind scheme, \emph{i.e.} its value is computed using Mualem-Van Genuchten model~\eqref{eq:dimless_MvG_K} for $h=h_{u(i,j)}^{\prime t+1}$, where $u_{i,j}=i$ if the flow is going from cell $i$ to $j$, and $u_{i,j}=j$ otherwise.

\item The change of both $\theta$ and $K$ is estimated using the first two terms of the Taylor series. If the time step is short enough, the algorithm converges to $h_i^{t+1}$ satisfying the continuity condition. Equation~\eqref{eq:Richard_discretisied} is linear in $h_i^{\prime t+1}$ for all $i$, and therefore, it can be solved using standard methods for linear algebraic equations.

\item After each iteration of groundwater flow, a number of iterations of overland flow is performed. \edit{In order to guarantee numeric stability, the Courant number defined as
\begin{equation}
    C=\frac{u_{i,j}\Delta t}{\|\mathbf{r}_{i\rightarrow j}\|}\qquad\text{with}\qquad u_{i,j}=K_{i,j}'\frac{h_j^{t+1}+z_j-h_i^{t+1}-z_i}{\|\mathbf{r}_{i\rightarrow j}\|},
\end{equation}
where $u_{i,j}$ represents the flow speed between cell $i$ and $j$, should be lower than $1$ for all pairs of computational cells. In order to achieve this, an adaptive time stepping is used to keep the Courant number at a given threshold value; however, additionally a minimum number of iterations is also preset to maintain high accuracy.}

\end{enumerate}

After each groundwater solver step for each cell with a face on the land surface, we compute the total volume of the water (surface and subsurface) divided by the total area of the top/bottom face ($\Delta x \Delta y$). Let us denote this ratio as $f_{i,j}$, where $i$ and $j$ are the given cell's indices. In each iteration of the surface solver, $h_s$ (and $h_c$ in 3D simulations) is computed for each cell as $h_{i,j}=f_{i,j}-f^\mathrm{max}_{i,j}$ for $f_{i,j}>f^\mathrm{max}_{i,j}$ and $h_{i,j}=0$ otherwise. Here, $f^\mathrm{max}_{i,j}$ is the maximum possible value of $f_{i,j}$, corresponding to a saturated cell with $h_g=0$. Then $f_{i,j}$ is updated using the following explicit scheme for flow given by the discretized form of the 2D Saint Venant equation~\eqref{eq:dimless_overland}:
\begin{equation}
    \begin{split}
        f_{i,j}^{t+1} = f_{i,j}^t + \frac{1}{\Delta x}\left(h_{i+1,j}^{5/3}-h_{i,j}^{5/3}\right)\sqrt{S_x} \frac{\Delta t}{n_s} + R_\mathrm{eff}.
    \end{split}
\end{equation}

After the last iteration of the overland flow scheme, $f_{i,j}$ values are used to calculate the pressure head $h_g$ in the subsurface cells bordering the land surface, after which the next time step for subsurface flow is computed. \edit{After each step we evaluate the output flow. In the case of the 3D model, we calculate the river flow at the outlet $Q(t)=q_c(\hat y=0, t)$, and in the case of the 2D model, we calculate the river inflow $\qin(t)$. These functions can then be presented in the form of a hydrograph.}

In addition to the above time-dependent solver, a steady-state solver was also implemented. It is based on the discretisation in~\eqref{eq:Richard_discretisied}, where the left-hand side (temporal term) is equal to $0$. The overland flow is included as an additional flow component between the surface cells and is solved simultaneously with the Richards equation.

The implementation described in this section was verified by successfully replicating the benchmark results by \cite{sulis2010comparison} obtained for a hillslope using the ParFlow integrated catchment model\edit{, and by \cite{maxwell2014surface} for the V-shaped catchment using the PAWS model (and other coupled surface-subsurface models).} The results of this comparison are presented in \cref{app:code_verification}.

\subsection{Example three-dimensional solution}
\label{sec:3D_time_dependent_solution}

\begin{table}
    \centering
    \begin{tabular}{ccc}
        \textsc{parameter} & \textsc{default value} & \textsc{parameter range} \\
        \hline
        $K_s\;[\mathrm{ms^{-1}}]$ & $1\cdot 10^{-5}$ & $10^{-6}-10^{-4}$ \\
        $L_{\hat x}\;[\mathrm{m}]$ & $6.16\cdot 10^2$ & $10^2-10^3$ \\
        $L_z\;[\mathrm{m}]$ & $6.84\cdot 10^2$ & $10^1-10^3$ \\
        $S_x\;[-]$ & $7.5\cdot 10^{-2}$ & $10^{-2}-10^{-1}$ \\
        $r\;[\mathrm{ms^{-1}}]$ & $2.36\cdot 10^{-7}$ & $3\cdot 10^{-8}-3\cdot 10^{-6}$ \\
        $r_0\;[\mathrm{ms^{-1}}]$ & $2.95\cdot 10^{-8}$ & $10^{-9}-10^{-7}$ \\
        $n_s\;[\mathrm{ms^{-1/3}}]$ & $5.1\cdot 10^{-2}$ & $10^{-2}-10^{-1}$ \\
        $\alpha_\mathrm{MvG}\;[\mathrm{m^{-1}}]$ & $3.7$ & $10^0-10^1$
    \end{tabular}
    \quad
    \begin{tabular}{cc}
        \textsc{parameter} & \textsc{value} \\
        \hline
        $L_y\;[\mathrm{m}]$ & \edit{$945$} \\
        $w\;[\mathrm{m}]$ & $5$ \\
        $n_s\;[\mathrm{ms^{-1/3}}]$ & $0.1$ \\
        $\theta_s\;[-]$ & $0.488$ \\
        $\theta_r\;[-]$ & $0$ \\
        $n\;[-]$ & $1.19$ \vspace{9.5mm}
    \end{tabular}
    \newline
    \vspace{2mm}
    \caption{Columns on right present the default values and ranges of the parameters used to perform a sensitivity analysis. Columns on the left present parameters, which were not varied during the sensitivity analysis. \edit{The length of the catchment was selected to be $L_y^\text{trib}$ estimated from Part 1, which represents the typical distance between the main tributaries of the river.}}
    \label{tab:simulation_settings}
\end{table}

\noindent Before proceeding to the quantitative analysis, we dedicate this section to a qualitative discussion of the general properties of a typical solution for the presented model. Let us consider a scenario of an intensive rainfall over the V-shaped catchment that initially remains in equilibrium with the mean precipitation characterising the given region. In the experiments, we find a steady-state solution for a mean precipitation $r_0$, which is set as the initial condition. We then simulate the reaction of the system to a constant precipitation $r>r_0$, and analyse the resulting river flow.

In the simulations, unless stated otherwise, all catchment parameters will be set to the typical values characterising UK catchments extracted in Part 1, which are summarised in \cref{tab:simulation_settings}. \edit{For the catchment length, $L_y$, we took the average length of the river between consecutive tributaries, $L_y^\text{trib}=945$ m, since at this scale, the drainage network no longer exhibits its fractal-like finger pattern. This way, we can treat our benchmark model as a representation of a single first-order catchment \citep{dietrich1993channel}, or a first/higher-order stream, forming the base element of more complex drainage networks \citep{strahler1957quantitative}.} The simulation results, covering a 24-hour rainfall, are collected and presented in \cref{fig:3D_solution_infographics}.

\afterpage{
\begin{figure}
    \vspace{3mm}    \includegraphics{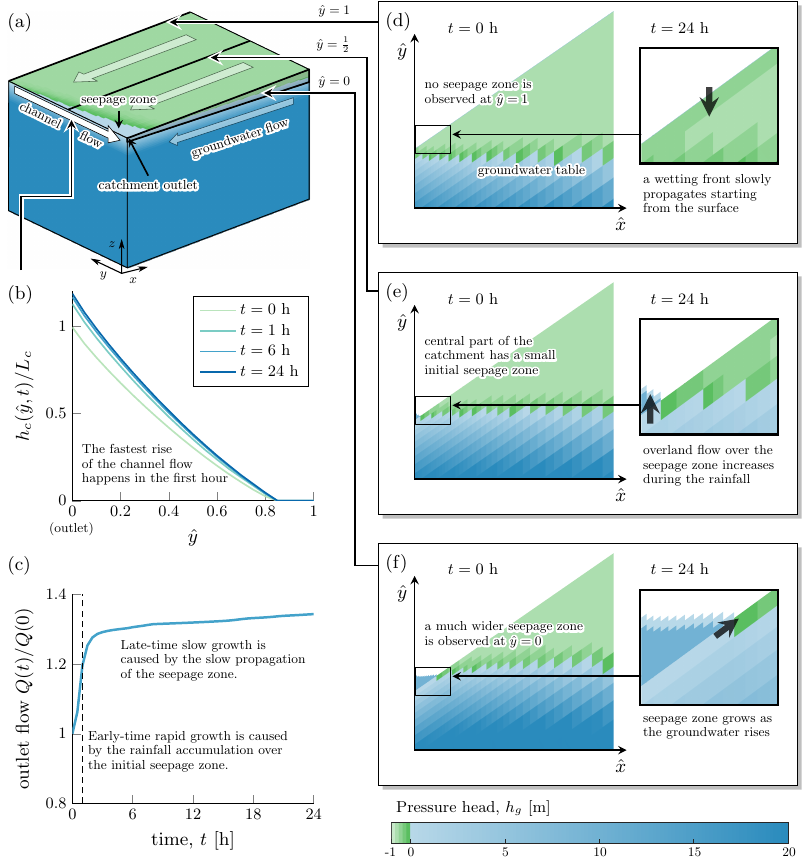}
    \vspace{-3mm}
    \captionsetup{type=figure}
    \caption{The illustration summarises the key properties of the 3D solution obtained for the default values of parameters from \cref{tab:simulation_settings}. \edit{Figure (a) depicts the initial condition (steady state for mean rainfall $r_0=2.95\cdot 10^{-8}$ m/s). During the subsequent rainfall $r=2.36\cdot 10^{-7}$ m/s, the water level in the channel rises as presented in (b), causing the flow at the river outlet to increase, as depicted by the hydrograph (c). We observe slightly different solution profiles depending on the location along the $\hat y$ axis -- their main features are outlined in cross-sections (d)-(e). The surface water height $h_s$ was magnified 5000 times, to make it visible.}}
    \label{fig:3D_solution_infographics}
\end{figure}
}

Initially, the system remains in a steady state, in which the pressure head $h_g$ increases with depth following an approximately hydrostatic profile (\cref{fig:3D_solution_infographics}a). The interesting dynamics responsible for generating the flow occurs near the surface ($\hat{z}=1$).

The surface water is present near the channel and extends further away from it for lower $\hat{y}$ values (\cref{fig:3D_solution_infographics}d-f). However, around $\hat y=1$, we do not observe surface water at all. This is caused by a non-zero gradient $S_y$ along the $\hat{y}$ direction, forcing the groundwater flow in that direction. We will refer to the zone in which the groundwater reaches the surface and forms an overland flow as \textit{the seepage zone} (see \cref{fig:2D_BCs}).

\edit{Two distinct effects are observed when the intensive rainfall starts. Firstly, the rainfall over the seepage zone starts to accumulate, causing the surface water to quickly rise (as highlighted in \cref{fig:3D_solution_infographics}e). Increased overland flow, leads to a rapid rise of the channel water and resulting outflow from the catchment within the first hour (\cref{fig:3D_solution_infographics}b-c).}

\edit{Secondly, the rainfall outside the seepage zone starts to infiltrate the unsaturated section of the soil, forming a characteristic wetting front (as highlighted in \cref{fig:3D_solution_infographics}d). After the infiltrating water reaches the groundwater, its level starts to rise. The rising groundwater eventually reaches the surface, causing the growth of the seepage zone (as in \cref{fig:3D_solution_infographics}f), increasing the area from which overland flow reaches the river.}

An essential observation for this time-dependent solution is that the characteristic timescale of overland flow dynamics is much shorter than the characteristic time of groundwater flow (their ratio is given by the dimensionless parameter $\tau_s\approx 2.8\cdot 10^{-4}$). This timescale separation is reflected in the shape of the hydrograph in \cref{fig:3D_solution_infographics}c, which shows the dependence between river flow at the outlet $Q(t)=q_c(\hat y,t)$ and time $t$.

A multiscale behaviour can be observed, with an early-time fast rise of total flow dominated by a rising overland flow fed by the rainfall over the seepage zone, and a late-time slow rise of total flow caused by rising groundwater and the resulting slow expansion of the seepage zone over time. This observation allows us both to understand the importance of model parameters (\cref{sec:sensitivity_analysis}) and to further simplify the problem in Part 3. \edit{Note that for the typical parameters studied in this work, the outlet flow will continue rising, eventually reaching a new steady state with $\lim_{t\rightarrow\infty}Q(t)=rA$. However, in case of the default simulation settings discussed above, it requires months of constant high rainfall for the solution to approach the new steady state. Thus, this effect is not observable over the typical day-long scales of the presented simulations.}

\section{Verification of 3D to 2D reduction}
\label{sec:3D_to_2D_verification}

\noindent The time-dependent simulations presented in the previous section demonstrate some of the three-dimensional features that are visible in the simulations. In this section, we further investigate these features, and show how they depend on two model parameters characterising the catchment geometry along the $\hat{y}$ direction, namely the catchment length $L_y$ and the slope parallel to the channel $S_y$. Alternatively, in terms of the nondimensional quantities, this corresponds to $\betazy$ and $\epsilon$. 

\subsection{Three-dimensional features of steady state solution}
\label{sec:3D_steady_states}

\begin{figure}
    \centering
    \includegraphics[trim=0 0 30 0, clip]{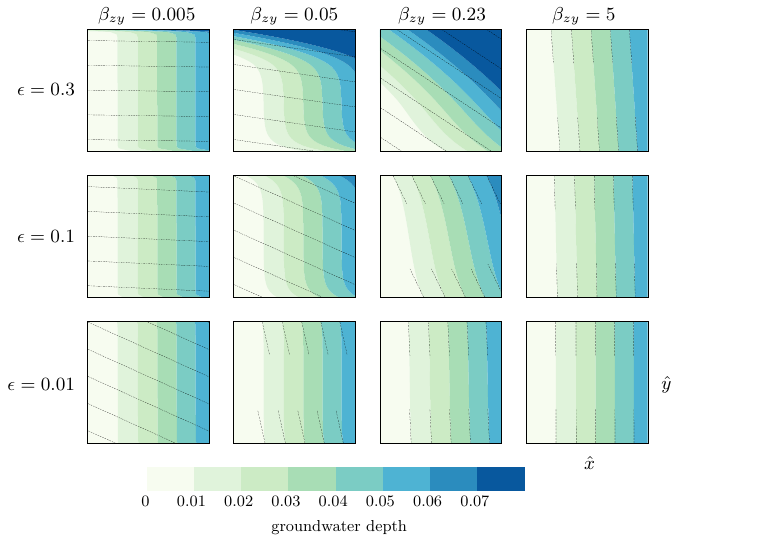}
    \caption{Groundwater table depth in steady states obtained for varying catchment length $L_y=\betazy^{-1}L_z$ and slope $S_y=\epsilon S_x$. Dashed lines represent lines of constant elevation. The entry with $\epsilon = 0.1$ and $\beta_{zy} = 0$ is the same figure as presented in \cref{fig:boundary_layer_diagram}.}
    \label{fig:3D_vs_2D_steady_states}
\end{figure}

\noindent In order to develop a better understanding of the three-dimensional effects, we performed a series of numerical experiments in which we found a steady state for varying values of $L_y$ and $S_y$, while keeping other parameters constant with the values provided in \cref{tab:simulation_settings}. The groundwater table shape corresponding to the selected steady states is presented in \cref{fig:3D_vs_2D_steady_states}.

\begin{figure}
    \centering
    \includegraphics{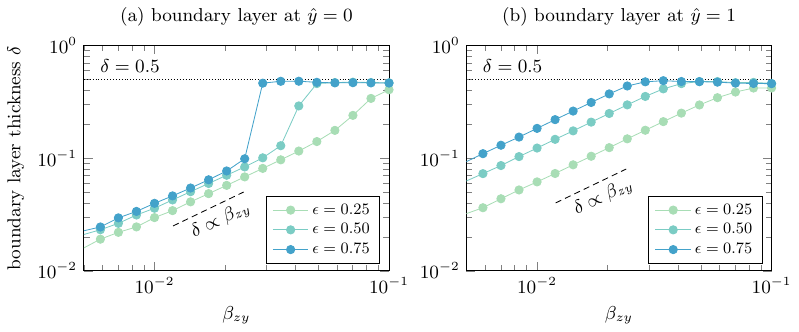}
    \caption{Boundary layer thickness at $\hat{y}=0$ (a) and $\hat{y}=1$ (b) as a function of $\betazy$. The boundary thickness was measured based on the groundwater depth profile along $\hat{x}\approx 0.46$. The boundary was defined as $\hat{y}$, for which the groundwater depth $H(0.46,\hat{y})$ is further than $\pm 5\%$ from the groundwater depth evaluated in the middle of the domain, $H(0.46, \hat{y}=0.5)$. For small $\betazy$ values, the boundary width follows $\delta\propto\betazy$ scaling. \edit{As $\betazy$ increases, $\delta$ reaches $0.5$, for which the boundary condition affects effectively the entire domain.}}
    \label{fig:boundary_layer_thickness}
    \vspace{-1mm}
\end{figure}

In all cases we observe that the solution becomes less $\hat{y}$-dependent as $\epsilon\rightarrow 0$, as expected from \cref{sec:reduced_models}. However, the dependence on $\beta_{zy}=L_z/L_y$ is more complex. The phase space can be divided into three regions:

\begin{enumerate}[label={(\roman*)},leftmargin=*, align = left, labelsep=\parindent, topsep=3pt, itemsep=2pt,itemindent=0pt ]
    \item When $L_y\ll \frac{L_{\hat x}}{\epsilon}$, the lines of constant elevation are approximately perpendicular to the hillslope (\emph{e.g.} $\epsilon=0.1$, $L_y=100$). As shown in \cref{sec:reduced_models}, in such a case, the leading-order (two-dimensional) solution of the governing equations for small $\epsilon$ also satisfies the boundary conditions in the leading order. In this case, we observe that the leading-order solution follows the lines of constant elevation over the entire domain.
    
    \item When $L_y \gg \frac{L_{\hat x}}{\epsilon}$, the lines of constant elevation are approximately parallel to the hillslope (\emph{e.g.} $\epsilon=0.1$, $L_y=10^6$). In such a case, the leading-order (2D) solution for the governing equations does not satisfy the flow boundary condition at $\hat y=0$ and $\hat y=1$. As a consequence, a boundary layer is developed near these two boundaries, in which the lines of constant groundwater table depth become parallel to lines of constant elevation, while in the outer solution they become perpendicular to the hillslope. The thickness of these boundary layers $\delta$ decreases inversely proportional to $L_y$ (see \cref{fig:boundary_layer_thickness}), which is consistent with the theoretical scaling derived in \cref{sec:limit_beta}.
    
    \item In the intermediate region, when $L_y = O\left(\frac{L_{\hat x}}{\epsilon}\right)$ and $\delta=O(1)$ (\emph{e.g.} $\epsilon=0.1$ and $L_y=3000$), the leading-order solution does not satisfy the boundary conditions, and the 'boundary layer' thickness becomes large enough to impact the solution over a major part or even effectively the entire domain. In such cases, the solution does not seem to satisfy the two-dimensional approximation unless $\epsilon$ is small enough.
\end{enumerate}

\subsection{Analysis of 3D to 2D reduction error}

\noindent Here, we follow the qualitative discussion from the previous section by quantifying the difference between the full solution and its two-dimensional approximation.

In \cref{sec:reduced_models}, we argued that the solution for the three-dimensional problem can be represented as $h_g(\hat x,\hat y,\hat z, t)=h_{g,0}(\hat x,\hat y, t)+\epsilon h_{g,1}(\hat x,\hat y,\hat z, t)+O(\epsilon^2)$, where $h_{g,0}$ is a two-dimensional solution for $\epsilon=0$. In this section, we verify this theoretical result numerically in the case of a steady-state solution, to which the same argument also applies.

\begin{figure}
    \centering
    \includegraphics{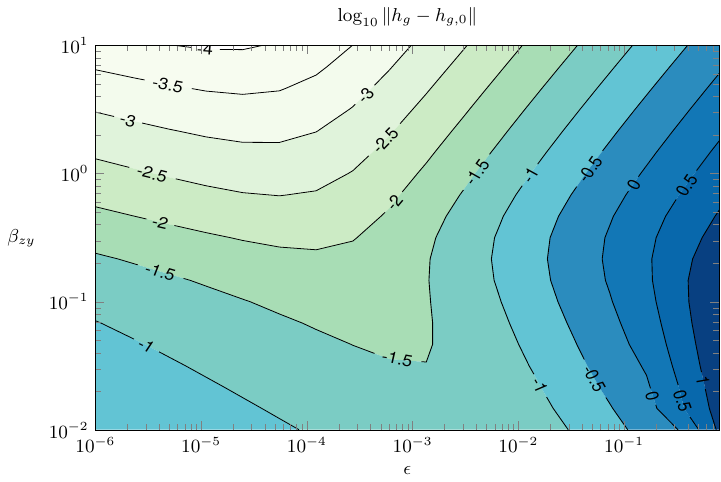}
    \caption{The mean absolute difference between the full three-dimensional solution and its two-dimensional approximation for small $\epsilon$ as a function of $\epsilon$ and $\beta_{zy}$.}
    \label{fig:3D_vs_2D_countour_plot}
\end{figure}

Finally, we estimated the mean absolute difference between $h_g$ and $h_{g,0}$ by averaging its values for all computational cells weighted by their volume. The dependence of this mean error on $L_y$ and $\epsilon$ is presented in \cref{fig:3D_vs_2D_countour_plot}. It confirms the asymptotic analysis from \cref{sec:reduced_models} and the qualitative observations from \cref{sec:3D_steady_states}. Firstly, it confirms that the error of the two-dimensional approximation increases proportionally to $\epsilon$ \edit{for large values of $\epsilon$. However for small values of $\epsilon$, the error increases, because the effect of y-dependent water height at the channel is no longer negligible (see \cref{sec:asymptotic_expansion_for_small_eps}).} Secondly, it shows that the error is small for very small values of $L_y$ (2D solution is satisfied everywhere) and very large $L_y$ values (2D solution is satisfied everywhere apart from a thin boundary layer at $\hat{y}=0$ and $\hat{y}=1$), but the error is the highest for intermediate values (here around $L_y=3000$).

\section{Impact of physical parameters on the 2D model}
\label{sec:parameters_impact}

\noindent Following \cref{sec:reduced_models}, the inflow to the river in our benchmark scenario for $S_y\ll S_x$ can be approximated by a two-dimensional model. In this section, we use the numerical procedure described in \cref{sec:numerical_methods} to quantify the impact of model parameters on the peak flows observed after an intensive rainfall and link them to the key physical processes accounting for the flow generation.

\subsection{Structure of typical hydrographs}

\noindent \edit{In this section, we examine, in more detail, the hydrographs that correspond to two protypical simulations. Under many sensible parameter choices, we have observed that many flow experiments can be roughly described into these two prototypical classes.  Although this is only described qualitatively in this work, we shall justify it more rigorously using the reduced model of Part 3.} 

\edit{For these simulations, we use the same parameter values as in \cref{sec:3D_time_dependent_solution}, with a catchment initially remaining in a steady state with rainfall $r_0$. The rainfall then rises to $r>r_0$ at $t=0$. Additionally, we set $S_y=0$ to reduce the problem dimension.}

\begin{figure}
    \centering
    \includegraphics[trim={1.6cm 0 2.5cm 0},clip,width=0.86\linewidth]{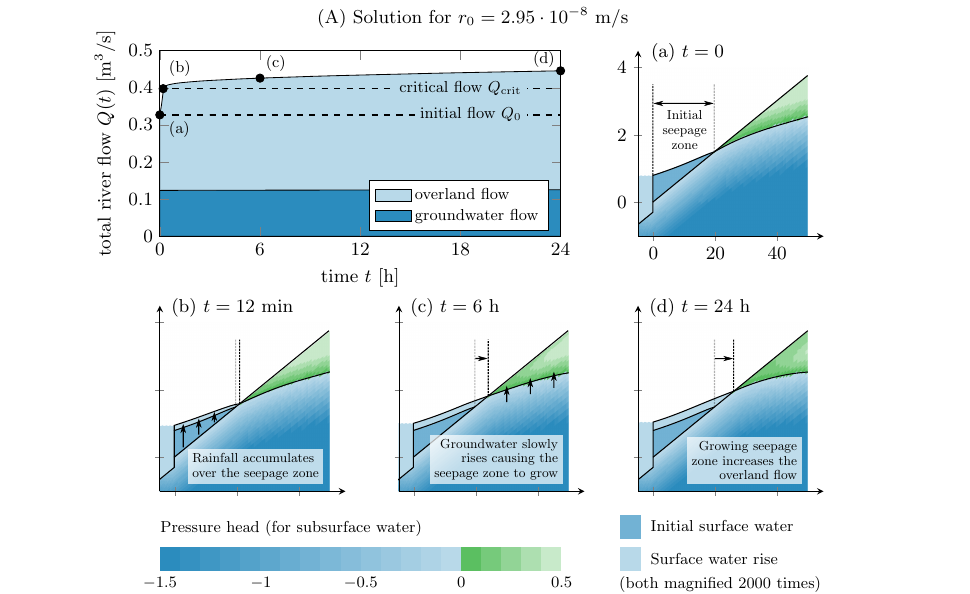}
    \caption{Numerical solution of 2D model for $r_0=2.95\cdot 10^{-8}$~ms$^{-1}$ with an initial seepage zone. All other parameters were set to the default values presented in \cref{tab:simulation_settings}.}
    \label{fig:Experiment_A}
\end{figure}

\begin{figure}
    \centering
    \includegraphics[trim={1.6cm 0 2.5cm 0},clip,width=0.86\linewidth]{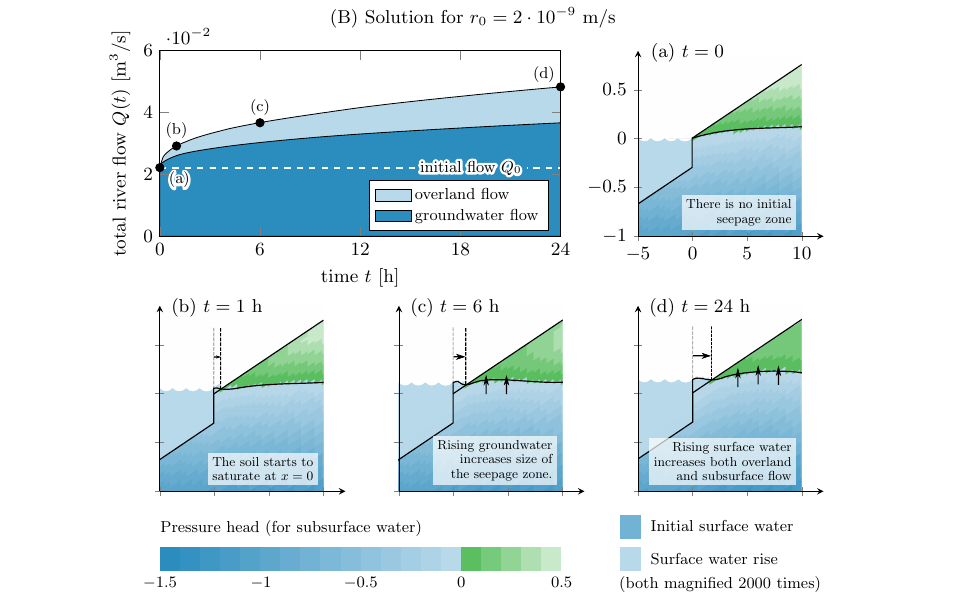}
    \caption{Numerical solution of 2D model for $r_0=2\cdot 10^{-9}$~ms$^{-1}$ without an initial seepage zone. All other parameters were set to the default values presented in \cref{tab:simulation_settings}.}
    \label{fig:Experiment_B}
\end{figure}

The numerical simulations are based on two experiments differentiated by their $r_0$ values: Experiment (A) with $r_0 = 2.95 \cdot 10^{-8}$ m/s; and Experiment (B) with $r_0 = 2 \cdot 10^{-9}$ m/s. The two corresponding hydrographs, $Q(t)$ vs $t$, are presented in the top-left inserts of \cref{fig:Experiment_A} and
\cref{fig:Experiment_B}. For each hydrograph, solutions are presented at four times and given in the insets (a) through (d). In the insets, areas shaded blue represent the saturated groundwater zone (with $h_g>0$), while shaded green areas represent the unsaturated zone. Surface water height was magnified $2000$ times, and its initial height was highlighted in a darker blue. Only a small part of the catchment near the river is presented.

The main difference between these two hydrographs is the existence of surface water in the initial condition. We observe that in the case presented in \cref{fig:Experiment_A}, the groundwater flow is not sufficient to transfer rainwater to the channel, and a fraction of the catchment area (namely the seepage zone) is initially covered with surface water. In contrast, in \cref{fig:Experiment_B}, initially there is no overland flow, \emph{i.e.} the groundwater never reaches the surface (except for the channel boundary). There is a significant qualitative difference between these two cases.

\subsubsection{Experiment (A): a case with an initial seepage zone}

\noindent \edit{In Experiment (A), we observe that the hydrograph can be roughly divided into two phases. We propose that in these two phases, the flow increase is determined by different physical mechanisms (similar to the three-dimensional case presented in \cref{fig:3D_solution_infographics}).}

\edit{During an early time phase (roughly first $12$ minutes), we observe a significant rise in total flow reaching the river. This corresponds to evolution between states (a) and (b) in fig. \ref{fig:Experiment_A}. We may interpret this early-time rise as a result of rainfall accumulating over the seepage zone, enhancing the already existing overland flow. This causes the river flow to rise by the rainfall excess over the initial seepage zone $(r-r_0)A_s$, where $A_s$ is the initial area of the seepage zone (which can be measured in the simulation).} 

\edit{As a result, in a short time, we define the flow
\begin{equation}
    \label{eq:saturation_flow}
    Q_\mathrm{crit} \equiv r_0A+(r-r_0)A_s.
\end{equation}
The above quantity we shall refer to as the \textit{critical flow}. Here, $r_0A$ represents the initial flow, and $A=L_xL_y$ is the catchment's area. In fact, the dashed line plotted in the hydrograph of figure\ref{fig:Experiment_A} is calculated via \eqref{eq:saturation_flow} and seems to coincide with the change in gradient of the hydrograph.}

\edit{At later times, we observe a slow growth of the total flow as a result of rising groundwater. Within this regime, the groundwater flow increases and the seepage zone slowly grows; consequently, there is an increased area over which an overland flow is generated. These effects cause the river flow, $Q(t)$, to exceed the critical flow $Q_\mathrm{crit}$. If $Q(t) \ll Q_\mathrm{crit}$, then the river flow is mostly caused by the early-time mechanism, while if $Q(t)\gg Q_\mathrm{crit}$, then the late-time mechanism dominates.} 

\edit{It should be noted that, here, we have introduced the intuition of the critical flow in \eqref{eq:saturation_flow} as a way to better interpret the numerical results. Shortly in \cref{sec:sensitivity_analysis}, we will justify based on sensitivity analysis of the model that many numerical solutions in the phase space do exhibit this behaviour (saturation to the critical flow). Moreover, in Part 3 of our work, we will derive $Q_\mathrm{crit}$ in a more rigorous way based on the asymptotic analysis based on a shallow-water approximation. For this case, the analogue to \eqref{eq:saturation_flow} will be developed asymptotically.}

\subsubsection{Experiment (B): case with no initial seepage zone}

\noindent In Experiment (B), there is no initial seepage zone. If the rainfall, $r$, is smaller than a certain value (dependent on soil geometry and properties around the channel), we may observe a slow rise in the groundwater table gradient around $x=0$, leading to an increase in the groundwater flow. If the rainfall is higher than this threshold value (as in the case presented in \cref{fig:Experiment_A}), then the gradient of the groundwater table eventually reaches the elevation gradient.

For typical values of rainfall much higher than the threshold value, this initial phase is very short and, in practice, not noticeable in the presented hydrograph. After that moment, a seepage zone starts to grow, giving rise to the overland flow, which slowly increases as the saturation front propagates. This is similar to the late-time behaviour of the first hydrograph. Additionally, we observe a rise in the groundwater flow as a result of the growing pressure head in the groundwater around the stream forced by the rising groundwater table. In the first case, the rise of the groundwater table was taking place far from the channel (relatively to its dimension), and so its effect on the groundwater recharge to the channel is not observable. 

\subsection{Sensitivity analysis}
\label{sec:sensitivity_analysis}

\begin{figure}
    \centering
    \includegraphics{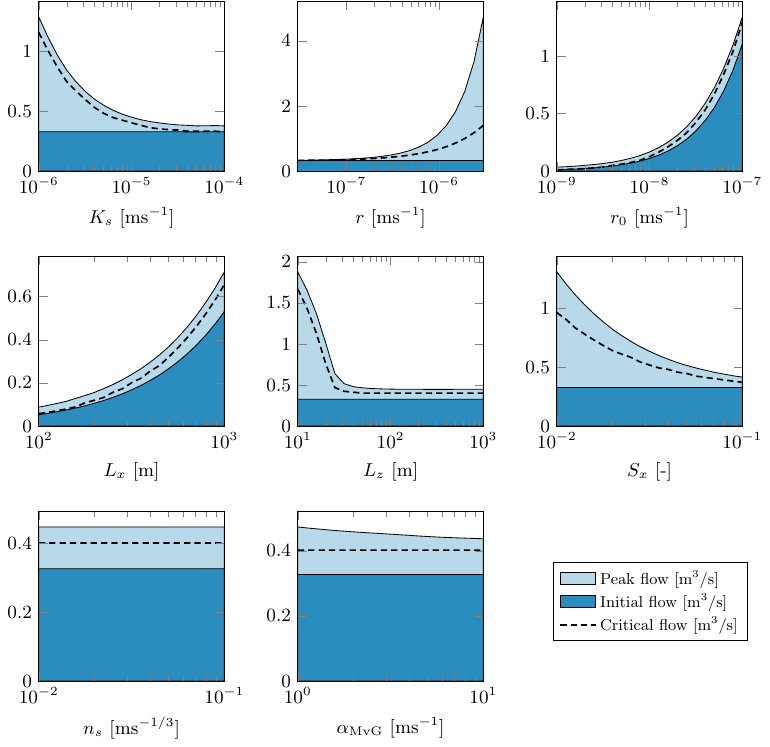}
    \caption{Results of the sensitivity analysis, showing the dependence of model parameters on the peak flow (light blue) and initial flow (dark blue). \edit{The y-axis on each figure represents the flow expressed in [$m^2/s$]. The peak flow is measured for a rainfall of duration of $t=24$ hours. The critical flow, represented with a dashed line, is defined by equation \eqref{eq:saturation_flow}.}}
    \label{fig:sensitivity_summary}
    \vspace{-1mm}
\end{figure}
 
\noindent In order to understand the relations between the described dynamics and model parameters, we conducted a sensitivity analysis. We chose eight physical parameters: catchment width $L_x$, aquifer depth $L_z$, elevation gradient along the hillslope $S_x$, hydraulic conductivity $K_s$, precipitation rates $r$ and $r_0$, Manning's constant $n_s$ and the $\alpha_\mathrm{MvG}$ parameter. Each parameter was varied within the range of its typical values presented in \cref{tab:simulation_settings} following \cite{paper1}, while keeping other parameters constant. In \cref{fig:sensitivity_summary}, we present the peak flow and its components after 24 hours, each as a function of the different parameter values. The critical flow calculated using~\eqref{eq:saturation_flow} is also shown on the graphs as a dashed line.

Based on this analysis and the investigation of the numerical solutions, the following conclusions can be drawn:
\begin{enumerate}[label={(\roman*)},leftmargin=*, align = left, labelsep=\parindent, topsep=3pt, itemsep=2pt,itemindent=0pt ]
    \item The critical flow generated by the precipitation accumulating over the initial seepage zone is a significant component of the peak flow; this description is consistent over the different model parameters.
    \item The size of this seepage zone depends on the difference between (i) the total precipitation in the initial condition, and (ii) the total groundwater flow. The former, (i), is a product of the precipitation rate, $r_0$, and the catchment area, $A=L_xL_y$, both of which are positively correlated with the seepage zone size. The latter, (ii), following Darcy's law, depends on hydraulic conductivity $K_s$, pressure gradient (dependent on slope $S_x$), and the aquifer depth $L_z$, all of which are negatively correlated with the size of the seepage zone.
    \item The precipitation rate, $r$, has a significant impact on both the critical flow as given by~\eqref{eq:saturation_flow}, and on the further growth of the overland flow. This is because it is responsible for the speed of groundwater rising and for surface water accumulation in the growing seepage zone.
    \item The speed of the seepage zone growth is slower for higher slope, $S_x$, values, since $S_x$ determines how deeply the groundwater table is located beneath the surface and how much rainwater it can absorb before reaching the surface. Also, $\alpha_\mathrm{MvG}$ has a small effect on the seepage zone growth, since it determines the soil saturation above the groundwater table. A higher $\alpha_\mathrm{MvG}$ causes the soil saturation to drop faster with height, allowing it to absorb more rainwater before it saturates. A similar effect is observed when varying other Mualem-van Genuchten model parameters ($\theta_S$, $\theta_R$, $n$). The impact of other model parameters on the hydrograph shape after reaching critical flow is very small.
    \item The soil depth, $L_z$, has a significant impact on the groundwater flow only for small values (comparable with the depth of the channel). Increasing $L_z$ above 30m has little impact on the solution, since the flow at such depths is insignificant.
    \item Manning's constant, $n_s$, seems to have almost no impact on the hydrograph. Its main contribution is in affecting the overland flow \edit{speed via the Manning's law \eqref{eq:2D_Manning}, and so it affects the characteristic timescale given by the $\tau_s$ parameter.} This timescale, however, is shorter than the duration of the simulated rainfall. The effect of the $n_s$ parameter can be significant if the rainfall duration is shorter than the time required to reach the critical flow (which is dependent on $n_s$). \edit{We will derive an analytical expression for this time in Part 3 of our work.} 
\end{enumerate}

\section{Discussion}
\label{sec:discussion}

\noindent The central question presented in our work is quite simple: \emph{What is the simplest three-dimensional model of coupled surface-subsurface flow on a hillslope?} 

Despite the fundamental nature of the above question, we have been surprised at the lack of mathematical and fluid dynamical research on issues of this nature in the literature. As mentioned throughout, we have been strongly motivated by the recent work of \cite{maxwell2014surface}, who designed benchmark scenarios for the purpose of comparing computational catchment models. Here, our philosophy has been more comprehensive in nature, and we are interested in the analytical and computational properties of the model rather than using it as a means to an end. Our benchmark involves several improvements over those proposed previously, allowing us to replicate hydrographs similar to the ones observed in real-world systems.

This work provides deeper insight into the mathematical structure of coupled surface-subsurface models. We extract and interpret nine key dimensionless parameters. As we show using asymptotic methods, under certain initial and geometric conditions ($S_y\ll S_x$, $L_y\ll L_x$ or $L_y\gg L_x$), the original formulation of the 3D model can be reduced to a 2D form. We then numerically investigated the shape and scale of the three-dimensional features, which subsequently allows us to quantify the error in the 3D-to-2D reduction.

Our sensitivity analysis of the key physical parameters reveals several interesting dependencies. As we demonstrate, the peak flows observed during sufficiently long rainfalls are usually caused by two mechanisms. First, there is an early-time rise due to surface water accumulating in the part of the catchment already saturated before the rainfall was initiated. Second, there is a late-time effect due to the slow propagation of the seepage zone. This two-scale behaviour can be rigorously justified based on asymptotic analysis of the governing equations. In the accompanying Part 3 of our work, we study the situation of aquifers with a depth much smaller than the catchment width (the \textit{shallow aquifer} scenario in \cref{fig:simplified_catchment}). There, we shall demonstrate that a shallow-water approximation allows us to derive analytical scaling laws for the hydrograph, and hence precise quantification of the peak flows mentioned above.

We note some potential consequences of our benchmark model for future research. The (relative) simplicity of our benchmark, and the clear isolation of properties such as peak flow formation and their parametric dependencies, means that the benchmark can be used in future studies for intermodel comparison. For example, data-based methods, such as conceptual and statistical models, may exhibit a different dependence on catchment properties. Then, by isolating the reasons for such discrepancies, we may better understand the limitations of different classes of models. This potentially leads to the development of more theoretically-justified models in the future, which may offer improvements in accuracy over a wider range of scenarios.

\mbox{}\par
\noindent \textbf{Acknowledgements.} We thank Sean Longfield (Environmental Agency) for many useful interactions and for motivating this work via the 7th Integrative Think Tank hosted by the Statistical and Applied Mathematics CDT at Bath (SAMBa). We also thank Thomas Kjeldsen (Bath), Tristan Pryer (Bath), and Rob Lamb (Lancaster/JBA Trust) for insightful discussions. We are indebted to the reviewers and the JFM editorial team---their comments and suggestions were instrumental in the final development of this paper. Piotr Morawiecki is supported by a scholarship from the EPSRC Centre for Doctoral Training in Statistical Applied Mathematics at Bath (SAMBa), under the project EP/S022945/1.

\mbox{}\par
\noindent \textbf{Declaration of Interests.} The authors report no conflict of interest.

\bibliographystyle{plainnat}
\bibliography{bibliography}

\begin{thebibliography}{62}
\providecommand{\natexlab}[1]{#1}
\providecommand{\url}[1]{\texttt{#1}}
\expandafter\ifx\csname urlstyle\endcsname\relax
  \providecommand{\doi}[1]{doi: #1}\else
  \providecommand{\doi}{doi: \begingroup \urlstyle{rm}\Url}\fi

\bibitem[Abbott et~al.(1986{\natexlab{a}})Abbott, Bathurst, Cunge, O'Connell,
  and Rasmussen]{abbott1986introduction1}
M.~B. Abbott, J.~C. Bathurst, J.~A. Cunge, P.~E. O'Connell, and J.~Rasmussen.
\newblock An introduction to the {European Hydrological System—Systeme
  Hydrologique Europeen}, "{SHE}", 1: History and philosophy of a
  physically-based, distributed modelling system.
\newblock \emph{J. Hydrol.}, 87\penalty0 (1-2):\penalty0 45--59,
  1986{\natexlab{a}}.

\bibitem[Abbott et~al.(1986{\natexlab{b}})Abbott, Bathurst, Cunge, O'Connell,
  and Rasmussen]{abbott1986introduction2}
M.~B. Abbott, J.~C. Bathurst, J.~A. Cunge, P.~E. O'Connell, and J.~Rasmussen.
\newblock An introduction to the {European Hydrological System—Systeme
  Hydrologique Europeen}, "{SHE}", 2: Structure of a physically-based,
  distributed modelling system.
\newblock \emph{J. Hydrol.}, 87\penalty0 (1-2):\penalty0 61--77,
  1986{\natexlab{b}}.

\bibitem[Akan(1985)]{akan1985similarity}
A.~O. Akan.
\newblock Similarity solution of overland flow on pervious surface.
\newblock \emph{J. Hydraul. Eng.}, 111\penalty0 (7):\penalty0 1057--1067, 1985.

\bibitem[Beven(2018)]{beven2018hypothesis}
K.~Beven.
\newblock On hypothesis testing in hydrology: {W}hy falsification of models is
  still a really good idea.
\newblock \emph{Wiley Interdiscip. Rev.: Water}, 5\penalty0 (3):\penalty0
  e1278, 2018.

\bibitem[Beven and Binley(1992)]{beven1992future}
K.~Beven and A.~Binley.
\newblock The future of distributed models: model calibration and uncertainty
  prediction.
\newblock \emph{Hydrol. Processes}, 6\penalty0 (3):\penalty0 279--298, 1992.

\bibitem[Beven and Germann(2013)]{beven2013macropores}
K.~Beven and P.~Germann.
\newblock Macropores and water flow in soils revisited.
\newblock \emph{Water Resour. Res.}, 49\penalty0 (6):\penalty0 3071--3092,
  2013.

\bibitem[Beven et~al.(1987)Beven, Calver, and Morris]{beven1987institute}
K.~Beven, A.~Calver, and E.~M. Morris.
\newblock The institute of hydrology distributed model.
\newblock Technical report, Institute of Hydrology, 1987.

\bibitem[Beven(2011)]{beven2011rainfall}
K.~J. Beven.
\newblock \emph{Rainfall-runoff modelling: the primer}.
\newblock John Wiley \& Sons, 2011.

\bibitem[Bixio et~al.(2000)Bixio, Orlandini, Paniconi, and
  Putti]{bixio2000physically}
A.~C. Bixio, S.~Orlandini, C.~Paniconi, and M.~Putti.
\newblock Physically-based distributed model for coupled surface runoff and
  subsurface flow simulation at the catchment scale.
\newblock \emph{Comp. Meth. Surf. Wat. Sys. Hydro.}, 2000.

\bibitem[Bouma(1981)]{bouma1981soil}
J.~Bouma.
\newblock Soil morphology and preferential flow along macropores.
\newblock \emph{Agric. Water Manage.}, 3\penalty0 (4):\penalty0 235--250, 1981.

\bibitem[Brunner and Simmons(2012)]{brunner2012hydrogeosphere}
P.~Brunner and C.~T. Simmons.
\newblock Hydrogeosphere: a fully integrated, physically based hydrological
  model.
\newblock \emph{Groundwater}, 50\penalty0 (2):\penalty0 170--176, 2012.

\bibitem[Brutsaert and Nieber(1977)]{brutsaert1977regionalized}
W.~Brutsaert and J.~L. Nieber.
\newblock Regionalized drought flow hydrographs from a mature glaciated
  plateau.
\newblock \emph{Water Resour. Res}, 13\penalty0 (3):\penalty0 637--643, 1977.

\bibitem[Buckingham(1914)]{buckingham1914physically}
E.~Buckingham.
\newblock On physically similar systems; illustrations of the use of
  dimensional equations.
\newblock \emph{Phys. Rev.}, 4\penalty0 (4):\penalty0 345, 1914.

\bibitem[by~{DHI}(2017)]{MIKE2017manual}
{MIKE}~Powered by~{DHI}.
\newblock {MIKE SHE}. {V}olume 2: {R}eference guide, 2017.

\bibitem[Calver and Wood(1991)]{calver1991dimensionless}
A.~Calver and W.~L. Wood.
\newblock Dimensionless hillslope hydrology.
\newblock \emph{Proc. Inst. Civ. Eng.}, 91\penalty0 (3):\penalty0 593--603,
  1991.

\bibitem[Chaudhry(2007)]{chaudhry2007open}
M.~H. Chaudhry.
\newblock \emph{Open-channel flow}.
\newblock Springer Science \& Business Media, 2007.

\bibitem[Constantine(2015)]{constantine2015active}
P.~G. Constantine.
\newblock \emph{Active subspaces: Emerging ideas for dimension reduction in
  parameter studies}.
\newblock SIAM, 2015.

\bibitem[Crawford and Linsley(1966)]{crawford1966digital}
N.~H. Crawford and R.~K. Linsley.
\newblock Digital simulation in hydrology: {Stanford Watershed Model} 4.
\newblock Technical report, Dept. of Civil Engineering, Stanford Universit,
  1966.

\bibitem[Cui et~al.(2014)Cui, Welty, and Maxwell]{cui2014modeling}
Z.~Cui, C.~Welty, and R.~M. Maxwell.
\newblock Modeling nitrogen transport and transformation in aquifers using a
  particle-tracking approach.
\newblock \emph{Comput. Geosci.}, 70:\penalty0 1--14, 2014.

\bibitem[Dietrich and Dunne(1993)]{dietrich1993channel}
W.~E. Dietrich and T.~Dunne.
\newblock The channel head.
\newblock \emph{Channel network hydrology}, 799:\penalty0 175--219, 1993.

\bibitem[Dogan and Motz(2005)]{dogan2005saturated}
A.~Dogan and L.~H. Motz.
\newblock Saturated-unsaturated 3{D} groundwater model. i: {D}evelopment.
\newblock \emph{J. Hydrol. Eng.}, 10\penalty0 (6):\penalty0 492--504, 2005.

\bibitem[Donigian and Imhoff(2006)]{donigian2006history}
A.~S. Donigian and J.~Imhoff.
\newblock History and evolution of watershed modeling derived from the
  {Stanford Watershed Model}.
\newblock \emph{Watershed models}, pages 21--45, 2006.

\bibitem[Farthing and Ogden(2017)]{farthing2017numerical}
M.~W. Farthing and F.~L. Ogden.
\newblock Numerical solution of {R}ichards' equation: {A} review of advances
  and challenges.
\newblock \emph{Soil Sci. Soc. Am. J.}, 81\penalty0 (6):\penalty0 1257--1269,
  2017.

\bibitem[Gilbert et~al.(2016)Gilbert, Jefferson, Constantine, and
  Maxwell]{gilbert2016global}
J.~M. Gilbert, J.~L. Jefferson, P.~G. Constantine, and R.~M. Maxwell.
\newblock Global spatial sensitivity of runoff to subsurface permeability using
  the active subspace method.
\newblock \emph{Adv. Water Resour.}, 92:\penalty0 30--42, 2016.

\bibitem[Gu{\'e}rin et~al.(2019)Gu{\'e}rin, Devauchelle, Robert, Kitou,
  Dessert, Quiquerez, Allemand, and Lajeunesse]{guerin2019stream}
A.~Gu{\'e}rin, O.~Devauchelle, V.~Robert, T.~Kitou, C.~Dessert, A.~Quiquerez,
  P.~Allemand, and E.~Lajeunesse.
\newblock Stream-discharge surges generated by groundwater flow.
\newblock \emph{Geophys. Res. Lett.}, 46\penalty0 (13):\penalty0 7447--7455,
  2019.

\bibitem[Gupta et~al.(2006)Gupta, Beven, and Wagener]{gupta2006model}
H.~V. Gupta, K.~J. Beven, and T.~Wagener.
\newblock Model calibration and uncertainty estimation.
\newblock \emph{Encyclopedia of hydrological sciences}, 2006.

\bibitem[Haverkamp et~al.(1998)Haverkamp, Parlange, Cuenca, Ross, and
  Steenhuis]{haverkamp1998scaling}
R.~Haverkamp, J.~Y. Parlange, R.~Cuenca, P.~J. Ross, and T.~S. Steenhuis.
\newblock Scaling of the {R}ichards equation and its application to watershed
  modeling.
\newblock \emph{Scale dependence and scale invariance in hydrology}, pages
  190--223, 1998.

\bibitem[Hutton et~al.(2016)Hutton, Wagener, Freer, Han, Duffy, and
  Arheimer]{hutton2016most}
C.~Hutton, T.~Wagener, J.~Freer, D.~Han, C.~Duffy, and B.~Arheimer.
\newblock Most computational hydrology is not reproducible, so is it really
  science?
\newblock \emph{Water Resour. Res.}, 52\penalty0 (10):\penalty0 7548--7555,
  2016.

\bibitem[Jefferson et~al.(2015)Jefferson, Gilbert, Constantine, and
  Maxwell]{jefferson2015active}
J.~L. Jefferson, J.~M. Gilbert, P.~G. Constantine, and R.~M. Maxwell.
\newblock Active subspaces for sensitivity analysis and dimension reduction of
  an integrated hydrologic model.
\newblock \emph{Comput. Geosci.}, 83:\penalty0 127--138, 2015.

\bibitem[Kirchner(2006)]{kirchner2006getting}
J.W. Kirchner.
\newblock Getting the right answers for the right reasons: Linking
  measurements, analyses, and models to advance the science of hydrology.
\newblock \emph{Water Resour. Res.}, 42\penalty0 (3), 2006.

\bibitem[Kirkby and Beven(1979)]{kirkby1979physically}
M.~J. Kirkby and K.~J. Beven.
\newblock A physically based, variable contributing area model of basin
  hydrology.
\newblock \emph{Hydrol. Sci. J.}, 24\penalty0 (1):\penalty0 43--69, 1979.

\bibitem[Kolditz et~al.(2012)Kolditz, Bauer, Bilke, B{\"o}ttcher, Delfs,
  Fischer, G{\"o}rke, Kalbacher, Kosakowski, McDermott,
  et~al.]{kolditz2012opengeosys}
O.~Kolditz, S.~Bauer, L.~Bilke, N.~B{\"o}ttcher, J.-O. Delfs, T.~Fischer, U.~J.
  G{\"o}rke, T.~Kalbacher, G.~Kosakowski, C.~I. McDermott, et~al.
\newblock Opengeosys: an open-source initiative for numerical simulation of
  thermo-hydro-mechanical/chemical ({THM/C}) processes in porous media.
\newblock \emph{Environ. Earth Sci.}, 67\penalty0 (2):\penalty0 589--599, 2012.

\bibitem[Kollet et~al.(2017)Kollet, Sulis, Maxwell, Paniconi, Putti, Bertoldi,
  Coon, Cordano, Endrizzi, Kikinzon, et~al.]{kollet2017integrated}
S.~Kollet, M.~Sulis, R.~M. Maxwell, C.~Paniconi, M.~Putti, G.~Bertoldi, E.~T.
  Coon, E.~Cordano, S.O Endrizzi, E.~Kikinzon, et~al.
\newblock The integrated hydrologic model intercomparison project, {IH-MIP2}:
  {A} second set of benchmark results to diagnose integrated hydrology and
  feedbacks.
\newblock \emph{Water Resour. Res.}, 53\penalty0 (1):\penalty0 867--890, 2017.

\bibitem[Kollet and Maxwell(2006)]{kollet2006integrated}
S.~J. Kollet and R.~M. Maxwell.
\newblock Integrated surface--groundwater flow modeling: {A} free-surface
  overland flow boundary condition in a parallel groundwater flow model.
\newblock \emph{Adv. Water Resour.}, 29\penalty0 (7):\penalty0 945--958, 2006.

\bibitem[Kollet et~al.(2009)Kollet, Cvijanovic, Sch{\"u}ttemeyer, Maxwell,
  Moene, and Bayer]{kollet2009influence}
S.~J. Kollet, I.~Cvijanovic, D.~Sch{\"u}ttemeyer, R.~M. Maxwell, A.~F. Moene,
  and P.~Bayer.
\newblock The influence of rain sensible heat and subsurface energy transport
  on the energy balance at the land surface.
\newblock \emph{Vadose Zone J.}, 8\penalty0 (4):\penalty0 846--857, 2009.

\bibitem[Liu et~al.(2004)Liu, Chen, Li, and Singh]{liu2004two}
Q.~Q. Liu, L.~Chen, J.~C. Li, and V.P. Singh.
\newblock Two-dimensional kinematic wave model of overland-flow.
\newblock \emph{J. Hydrol.}, 291\penalty0 (1-2):\penalty0 28--41, 2004.

\bibitem[Markovich et~al.(2016)Markovich, Maxwell, and
  Fogg]{markovich2016hydrogeological}
K.~H. Markovich, R.~M. Maxwell, and G.~E. Fogg.
\newblock Hydrogeological response to climate change in alpine hillslopes.
\newblock \emph{Hydrol. Processes}, 30\penalty0 (18):\penalty0 3126--3138,
  2016.

\bibitem[Maxwell et~al.(2009)Maxwell, Kollet, Smith, Woodward, Falgout,
  Ferguson, Baldwin, Bosl, Hornung, and Ashby]{maxwell2009parflow}
R.~M. Maxwell, S.~J. Kollet, S.~G. Smith, C.~S. Woodward, R.~D. Falgout, I.~M.
  Ferguson, C.~Baldwin, W.~J. Bosl, R.~Hornung, and S.~Ashby.
\newblock Par{F}low user’s manual.
\newblock \emph{International Ground Water Modeling Center Report GWMI},
  1\penalty0 (2009):\penalty0 129, 2009.

\bibitem[Maxwell et~al.(2014)Maxwell, Putti, Meyerhoff, Delfs, Ferguson,
  Ivanov, Kim, Kolditz, Kollet, Kumar, et~al.]{maxwell2014surface}
R.~M. Maxwell, M.~Putti, S.~Meyerhoff, J.-O. Delfs, I.~M. Ferguson, V.~Ivanov,
  J.~Kim, O.~Kolditz, S.~J. Kollet, M.~Kumar, et~al.
\newblock Surface-subsurface model intercomparison: A first set of benchmark
  results to diagnose integrated hydrology and feedbacks.
\newblock \emph{Water Resour. Res.}, 50\penalty0 (2):\penalty0 1531--1549,
  2014.

\bibitem[Meyerhoff and Maxwell(2011)]{meyerhoff2011quantifying}
S.~B. Meyerhoff and R.~M. Maxwell.
\newblock Quantifying the effects of subsurface heterogeneity on hillslope
  runoff using a stochastic approach.
\newblock \emph{Hydrogeol. J.}, 19\penalty0 (8):\penalty0 1515--1530, 2011.

\bibitem[Morawiecki(2022)]{github2}
P.~W. Morawiecki.
\newblock {GitHub} repository for {3D}, {2D} and {1D} benchmark catchment
  models.
\newblock \url{https://github.com/Piotr-Morawiecki/benchmark-catchment-model},
  2022.
\newblock Accessed: 2022-10-28.

\bibitem[Morawiecki and Trinh(2022)]{paper1}
P.~W. Morawiecki and P.~H. Trinh.
\newblock On the development and analysis of coupled surface-subsurface models
  of catchments. {P}art 1. {A}nalysis of dimensions and parameters for uk
  catchments.
\newblock \emph{T.B.C.}, 2022.

\bibitem[Neuzil and Tracy(1981)]{neuzil1981flow}
C.~E. Neuzil and J.~V. Tracy.
\newblock Flow through fractures.
\newblock \emph{Water Resour. Res.}, 17\penalty0 (1):\penalty0 191--199, 1981.

\bibitem[Peel and McMahon(2020)]{peel2020historical}
M.~C. Peel and T.~A. McMahon.
\newblock Historical development of rainfall-runoff modeling.
\newblock \emph{Wiley Interdiscip. Rev.: Water}, 7\penalty0 (5):\penalty0
  e1471, 2020.

\bibitem[Rihani et~al.(2015)Rihani, Chow, and Maxwell]{rihani2015isolating}
J.~F. Rihani, F.~K. Chow, and R.~M. Maxwell.
\newblock Isolating effects of terrain and soil moisture heterogeneity on the
  atmospheric boundary layer: {I}dealized simulations to diagnose
  land-atmosphere feedbacks.
\newblock \emph{J. Adv. Model. Earth Syst.}, 7\penalty0 (2):\penalty0 915--937,
  2015.

\bibitem[Sanford et~al.(1993)Sanford, Parlange, and
  Steenhuis]{sanford1993hillslope}
W.~E. Sanford, J.-Y. Parlange, and T.~S. Steenhuis.
\newblock Hillslope drainage with sudden drawdown: Closed form solution and
  laboratory experiments.
\newblock \emph{Water Resour. Res}, 29\penalty0 (7):\penalty0 2313--2321, 1993.

\bibitem[Schaake~Jr(1975)]{schaake1975surface}
J.~C. Schaake~Jr.
\newblock Surface waters.
\newblock \emph{Rev. Geophys.}, 13\penalty0 (3):\penalty0 445--451, 1975.

\bibitem[Shaw et~al.(2010)Shaw, Beven, Chappell, and Lamb]{shaw2015hydrology}
E.~Shaw, K.~Beven, N.~Chappell, and R.~Lamb.
\newblock \emph{Hydrology in practice}.
\newblock CRC press, 3 edition, 2010.

\bibitem[Shen and Phanikumar(2010)]{shen2010process}
C.~Shen and M.~S. Phanikumar.
\newblock A process-based, distributed hydrologic model based on a large-scale
  method for surface--subsurface coupling.
\newblock \emph{Adv. Water Resour.}, 33\penalty0 (12):\penalty0 1524--1541,
  2010.

\bibitem[Singh and Frevert(2003)]{singh2003watershed}
V.~P. Singh and D.~K. Frevert.
\newblock Watershed modeling.
\newblock In \emph{World Water \& Environmental Resources Congress 2003}, pages
  1--37, 2003.

\bibitem[Sitterson et~al.(2018)Sitterson, Knightes, Parmar, Wolfe, Avant, and
  Muche]{sitterson2018overview}
J.~Sitterson, C.~Knightes, R.~Parmar, K.~Wolfe, B.~Avant, and M.~Muche.
\newblock An overview of rainfall-runoff model types.
\newblock In \emph{Proceedings of 9th Int. Congr. Env. Mod. Soft}, 2018.

\bibitem[Sivapalan et~al.(1987)Sivapalan, Beven, and
  Wood]{sivapalan1987hydrologic}
M.~Sivapalan, K.~Beven, and E.~F. Wood.
\newblock On hydrologic similarity: 2. {A} scaled model of storm runoff
  production.
\newblock \emph{Water Resour. Res.}, 23\penalty0 (12):\penalty0 2266--2278,
  1987.

\bibitem[Strahler(1957)]{strahler1957quantitative}
A.~N. Strahler.
\newblock Quantitative analysis of watershed geomorphology.
\newblock \emph{Eos, Transactions American Geophysical Union}, 38\penalty0
  (6):\penalty0 913--920, 1957.

\bibitem[Sulis et~al.(2010)Sulis, Meyerhoff, Paniconi, Maxwell, Putti, and
  Kollet]{sulis2010comparison}
M.~Sulis, S.~B. Meyerhoff, C.~Paniconi, R.~M. Maxwell, M.~Putti, and S.~J.
  Kollet.
\newblock A comparison of two physics-based numerical models for simulating
  surface water--groundwater interactions.
\newblock \emph{Adv. Water Resour.}, 33\penalty0 (4):\penalty0 456--467, 2010.

\bibitem[Sulis et~al.(2017)Sulis, Williams, Shrestha, Diederich, Simmer,
  Kollet, and Maxwell]{sulis2017coupling}
M.~Sulis, J.~L. Williams, P.~Shrestha, M.~Diederich, C.~Simmer, S.~J. Kollet,
  and R.~M. Maxwell.
\newblock Coupling groundwater, vegetation, and atmospheric processes: {A}
  comparison of two integrated models.
\newblock \emph{J. Hydrometeorol.}, 18\penalty0 (5):\penalty0 1489--1511, 2017.

\bibitem[Sweetenham et~al.(2017)Sweetenham, Maxwell, and
  Santi]{sweetenham2017assessing}
M.~G. Sweetenham, R.~M. Maxwell, and P.~M. Santi.
\newblock Assessing the timing and magnitude of precipitation-induced seepage
  into tunnels bored through fractured rock.
\newblock \emph{Tunnelling Underground Space Technol.}, 65:\penalty0 62--75,
  2017.

\bibitem[Tayfur and Kavvas(1994)]{tayfur1994spatially}
G.~Tayfur and M.~L. Kavvas.
\newblock Spatially averaged conservation equations for interacting
  rill-interrill area overland flows.
\newblock \emph{J. Hydraul. Eng.}, 120\penalty0 (12):\penalty0 1426--1448,
  1994.

\bibitem[Van~Genuchten(1980)]{van1980closed}
M.~Th. Van~Genuchten.
\newblock A closed-form equation for predicting the hydraulic conductivity of
  unsaturated soils.
\newblock \emph{Soil Sci. Soc. Am. J.}, 44\penalty0 (5):\penalty0 892--898,
  1980.

\bibitem[Vieira(1983)]{vieira1983conditions}
J.~H.~Daluz Vieira.
\newblock Conditions governing the use of approximations for the {Saint-Venant}
  equations for shallow surface water flow.
\newblock \emph{J. Hydrol.}, 60\penalty0 (1-4):\penalty0 43--58, 1983.

\bibitem[Warrick and Hussen(1993)]{warrick1993scaling}
A.~W. Warrick and A.~A. Hussen.
\newblock Scaling of {R}ichards' equation for infiltration and drainage.
\newblock \emph{Soil Sci. Soc. Am. J.}, 57\penalty0 (1):\penalty0 15--18, 1993.

\bibitem[Warrick et~al.(1990)Warrick, Lomen, and Islas]{warrick1990analytical}
A.~W. Warrick, D.~O. Lomen, and A.~Islas.
\newblock An analytical solution to {R}ichards' equation for a draining soil
  profile.
\newblock \emph{Water Resour. Res.}, 26\penalty0 (2):\penalty0 253--258, 1990.

\bibitem[Weill et~al.(2009)Weill, Mouche, and Patin]{weill2009generalized}
S.~Weill, E.~Mouche, and J.~Patin.
\newblock A generalized {R}ichards equation for surface/subsurface flow
  modelling.
\newblock \emph{J. Hydrol.}, 366\penalty0 (1-4):\penalty0 9--20, 2009.

\end{thebibliography}

\appendix

\newpage
\section{List of symbols} \label{sec:listofsymbols}

\noindent For convenience, we provide a list of symbols in tables~\ref{tab:list1} and \ref{tab:list2}. The definitions of the dimensionless parameters is provided in \cref{app:list_of_parameters}.
\vspace{3mm}

{
\begin{tabular}{p{0.15\textwidth}p{0.17\textwidth}p{0.65\textwidth}}
    \toprule
    \textsc{group} & \textsc{symbol} & \textsc{description} \\
    \midrule
    Independent & $t$ & time \\
    variables & $x$, $y$, $z$ & catchment coordinates \\
    & $\hat{x}$, $\hat{y}$, $\hat{z}$ & tilted coordinates \\
    \midrule
    Groundwater & $h_g$ & pressure head \\
    flow & $h_{g,0}$, $h_{g,1}$ & terms of asymptotic expansion of $h_g$ \\
    & $K_s$ & saturated soil conductivity \\
    & $K_r$ & relative hydraulic conductivity \\
    & $\theta$ & volumetric water content \\
    & $\theta_s$ & residual water content \\
    & $\theta_r$ & saturated water content \\
    & $\alpha_\mathrm{MvG}$ & Mualem-van Genuchten model $\alpha$ parameter \\
    & $n$, $m$ & MvG model parameters quantifying pore size distribution \\
    \midrule
    Overland and & $h_s, h_c$ & water height on the land surface and in the channel \\
    channel flow & $h_{s,0}$, $h_{s,1}$ & terms of asymptotic expansion of $h_s$ \\
    & $\mathbf{q_g}$, $\mathbf{q_s}$, $q_c$ & groundwater, overland, and channel flow \\
    & $v_s$, $v_c$ & velocity of overland and channel flow \\
    & $\mathbf{S}_f$ & friction slope for the overland flow \\
    & $S_f^\mathrm{river}$ & friction slope for 1D channel flow \\
    & $R$ & rainfall rate \\
    & $ET$ & evapotranspiration rate \\
    & $I$ & surface water infiltration rate \\
    & $r_\mathrm{eff}$ & effective rainfall (defined as $R-ET$) \\
    & $n_s, n_c$ & Manning's $n$ coefficient for surface and channel  \\
    & $g$ & gravitational acceleration \\
    & $\qin$ & total surface and subsurface flow to the channel \\
    & $A$ & area of channel cross-section \\
    & $P$ & channel wetted perimeter \\
    \midrule
    Catchment & $L_x$, $L_{\hat x}$, $L_y$, $L_z$ & catchment/hillslope dimension along $x$, $\hat{x}$, $y$, and $z$ \\
    geometry & \edit{$H_\mathrm{surf}(\hat x, \hat y)$} & \edit{elevation of the land surface} \\
    & \edit{$\textbf{S}_0=-\nabla H_\mathrm{surf}$} & \edit{elevation gradient (slope)} \\
    & $S_x$, $S_y$ & slope measured along $\hat{x}$ and $y$ direction \\
    & $\phi$ & angle between the direction of the steepest descent and the $x$ direction \\
    & $w$ & channel width \\
    \midrule
    Scaling & $t_0$ & characteristic timescale \\
    factors & $L_s$, $L_c$ & characteristic overland and channel water height \\
    & $v_{s,0}^{\hat x}$, $v_{s,0}^{\hat y}$, $v_{c,0}$ & characteristic scale of overland and channel flow velocity \\
    \bottomrule
\end{tabular}
\captionsetup{type=table}
\captionof{table}{First list of symbols \label{tab:list1}}
}

\clearpage
{
\begin{tabular}{p{0.15\textwidth}p{0.15\textwidth}p{0.65\textwidth}} \footnotesize
\\ \toprule
    \textsc{group} & \textsc{symbol} & \textsc{description} \\
    \midrule
    Dimensionless &  $\alpha$ & dimensionless $\alpha_\mathrm{MvG}$ parameter \\
    parameters &  $\betazx$, $\betazy$ & aspect ratio of cross-section along the hillslope and channel \\
    & $\epsilon$ & ratio of $S_y$ to $S_x$ slope \\
    & $\gamma$ & aspect ratio of the stream's cross-section \\
    & $\lambda_s$, $\lambda_c$ & ratio between surface/channel characteristic water height and aquifer thickness $L_z$ \\
    & $\rho$ & dimensionless rainfall defined as $\rho=\frac{r}{K_s}$ \\
    & $\tau_s$, $\tau_c$ & ratio of overland/channel flow and groundwater timescale \\
    \midrule
    Numerical & $N_x$, $N_y$, $N_z$ & number of mesh cells along the $x$, $y$, and $z$ axes \\
    method & $\Delta t$ & time step duration \\
    & $V_i$ & volume of cell $i$ \\
    & $S_{i,j}$ & face area between cell $i$ and $j$\\
    & $\mathbf{r}_{i\rightarrow j}$ & vector from the centroid of cell $i$ to the centroid of cell $j$ \\
    & $\boldsymbol{\beta}$ & vector of $\beta$ parameters, $\boldsymbol{\beta}=(\betazx^2,\betazy^2,1)$ \\
    & $K_{i,j}'$ & hydraulic conductivity of the face between cell $i$ and $j$ \\
    & $u_{i,j}$ & function returning the index of the uplift cell ($i$ or $j$) \\
    & $f_{i,j}^t$ & water volume in surface cell $(i,j)$ divided by its base area \\
    & $\Delta x$, $\Delta y$ & extent of the cell in the $x$ and $y$ direction \\
\bottomrule
\end{tabular}
\captionsetup{type=table}
\captionof{table}{Second list of symbols \label{tab:list2}}
}

\newpage
\section{Governing equations is tilted coordinates}

\subsection{Dimensional form}
\label{app:governing_dim}

\noindent Here we write down the governing equations introduced in \cref{sec:Model_formulation} in $(\hat{x},\hat{y},\hat{z})$ coordinates as given by transformation~\eqref{eq:tilted_coordinates}.

The Richards equation~\eqref{eq:Richards} becomes
\begin{multline}
    \label{eq:Richards_xyz_form}
        \frac{1}{K_s}\frac{\d\theta}{\d h}\bigg\rvert_{h=h_g}\dt{h_g} = \left[1-\left(\frac{S_y}{S_x}\right)^2\right]\dxhat{} \left[K_r(h_g) \dxhat{h_g}\right] \\
        + \frac{S_y}{S_x} \dxhat{} \left[K_r(h_g) \left(\frac{S_y}{S_x}\dxhat{h_g}+\dyhat{h_g}\right)\right] + \dyhat{} \left[K_r(h_g) \left(\frac{S_y}{S_x}\dxhat{h_g}+\dyhat{h_g}\right)\right] \\
        + S_x\dxhat{} \left[K_r(h_g) \left(S_x\dxhat{h_g}+S_y\dyhat{h_g}+\dzhat{h_g}+1\right)\right] \\
        + S_y\dyhat{} \left[K_r(h_g) \left(S_x\dxhat{h_g}+S_y\dyhat{h_g}+\dzhat{h_g}+1\right)\right] \\
        + \dzhat{} \left[K_r(h_g) \left(S_x\dxhat{h_g}+S_y\dyhat{h_g}+\dzhat{h_g}+1\right)\right].
\end{multline}

The Saint Venant equation~\eqref{eq:St_Venant_overland}, together with \eqref{eq:2D_Manning} in transformed coordinates $\hat{x}$ and $\hat{y}$ becomes:
\begin{equation}
    \label{eq:2D_saint_Venant_xy_form}
    \dt{h_s}=\frac{1}{n_s}\dxhat{}\left(h_s^{5/3}\sqrt{S_x}\right)+R_\mathrm{eff},
\end{equation}
where $R_\mathrm{eff}=R-ET$ is the effective precipitation.

The channel flow is given by equations~\eqref{eq:St_Venant_channel} and~\eqref{eq:Manning_flow} combined, which for our simplified catchment give the last governing equation:
\begin{equation}
    \label{eq:St_Venant_rectangular_channel}
    \dt{h_c} = \frac{\qin}{w} - \frac{1}{n_c}\dyhat{}\left(h_c^{5/3}\sqrt{S_y}\right).
\end{equation}

All boundary conditions in the dimensional form are listed in \cref{sec:BC}.

\subsection{Dimensionless form}
\label{app:governing_dimless}

Now, we rewrite eqs~\eqref{eq:Richards_xyz_form}, \eqref{eq:2D_saint_Venant_xy_form}, \eqref{eq:St_Venant_rectangular_channel} using the dimensionless quantities introduced in \cref{sec:nondimensionalisation}. 
Here and henceforth, we shall drop the primes, and assume that all subsequent quantities are dimensionless. The dimensionless governing equations are as follows. First, the 3D Richards equation for pressure head, $h_g(\hat x,\hat y,\hat z)$:
\begin{multline}
    \label{eq:dimless_subsurface}
    \underbrace{\frac{\d\theta}{\d h}\bigg\rvert_{h=h_g}\dt{h_g}}_{\approx 1} =
    \underbrace{\dzhat{} \left[K_r(h_g) \left(\dzhat{h_g}+1\right)\right]}_{\approx 1}
    + \underbrace{\betazx S_x\dxhat{} \left[K_r(h_g) \left(2\dzhat{h_g}+1\right)\right]}_{\approx 10^{-1}} \\
    + \underbrace{\betazx^2 \left(1 + S_x^2\right)\dxhat{} \left[K_r(h_g) \dxhat{h_g}\right]}_{\approx 1}
    + \underbrace{\betazy^2 \left(1 + S_y^2\right) \dyhat{} \left[K_r(h_g)\dyhat{h_g}\right]}_{\approx 1\dagger} \\
    + \underbrace{2 \betazx \betazy\frac{S_y}{S_x}\left(1+S_x^2\right) \dxhat{} \left[K_r(h_g)\dyhat{h_g}\right]}_{\approx 10^{-1}\dagger}
    + \underbrace{\betazy S_y\dyhat{} \left[K_r(h_g) \left(2\dzhat{h_g}+1\right)\right]}_{\approx 10^{-2}\dagger}.
\end{multline}
The 2D Saint Venant equation for overland water height $h_s(\hat x,\hat y)$:
\begin{equation}
    \label{eq:dimless_overland}
    \underbrace{\tau_s\dt{h_s}}_{\approx 10^{-4}}=
    \underbrace{\dxhat{}\left(h_s^{5/3}\right)}_{\approx 1}+
    \underbrace{R_\mathrm{eff}-I}_{\approx 1}.
\end{equation}
Finally, the 1D Saint Venant equation for channel water height $h_s(\hat y)$:
\begin{equation}
    \label{eq:dimless_channel}
    \underbrace{\tau_c\dt{h_c}}_{\approx 10^{-3}} =
    \underbrace{\qin}_{\approx 1} - \underbrace{\dyhat{}\left(h_c^{5/3}\right)}_{\approx 1}.
\end{equation}

The definition of dimensionless parameters ($\betazx$, $\betazy$, $\tau_s$, $\tau_c$, $\gamma$), their interpretation, and estimated values are presented in \cref{app:list_of_parameters}. Numerical values under the equations represent the typical order of magnitude of parameters multiplying the given term. However, note that terms marked with "$\dagger$" symbol include the $\hat{y}$-derivative of the solution, which, as was discussed in \cref{sec:reduced_models}, is $\hat{y}$-independent in the leading order. The effect of the relative size of these terms is much smaller (by approximately one order of magnitude) than indicated by the provided values of prefactors.

In the above equations, the dimensionless $\theta(h)$ and $K_r(h)$ functions are given by:
\begin{subequations}
    \label{eq:MvG_dimless}
    \begin{align}
        \frac{\d\theta(h)}{\d h} &=
        \begin{cases}
            \frac{mn(\theta_s-\theta_r)}{h}\frac{(\alpha h)^n}{\left(1+\left(\alpha h\right)^n\right)^{m+1}} & h<0 \\
            0 & h\ge0
        \end{cases}, \\
        K_r(h) &=
        \begin{cases}
            \frac{\left(1-\left(\alpha h\right)^{n-1}\left(1+\left(\alpha h\right)^n\right)^{-m}\right)^2}{\left(1+\left(\alpha h\right)^n\right)^{m/2}} & h<0 \\
            1 & h\ge0
        \end{cases}, \label{eq:dimless_MvG_K}
    \end{align}
\end{subequations}
\noindent where $\alpha = \alpha_\mathrm{MvG} L_z$ is a dimensionless MvG $\alpha$ parameter.

Finally, as divided into the enumerations of \cref{sec:BC}, the non-dimensional boundary conditions are now as follows:
\begin{enumerate}[label={(\roman*)},leftmargin=*, align = left, labelsep=\parindent, topsep=3pt, itemsep=2pt,itemindent=0pt ]
\begin{subequations}
\item On the catchment boundary, $\Gamma_B$:
\begin{equation}
    \label{eq:bc_nondim_first}
    \mathbf{q_g}\cdot\mathbf{n} = 0, \quad \mathbf{q_s}\cdot\mathbf{n} = 0 \qquad \text{on $\Gamma_B$}.
\end{equation}
\item On the land surface, $\Gamma_s$:
\begin{equation} \label{eq:dimless_river_BC}
        h_s\Big\rvert_{\Gamma_s} =
        \begin{cases}
            0 & \quad \text{if } h_g<0\\
            \lambda_s^{-1} h_g  & \quad \text{if } h_g>0
        \end{cases} \quad \text{and} \quad
        \mathbf{q_g}\cdot\mathbf{n}\Big\rvert_{\Gamma_s} = \rho I.
\end{equation}
\item At the channel, $\Gamma_R$:
\begin{equation}
    \edit{h_g\Big\rvert_{\Gamma_R} = \lambda_c h_c - z.}
\end{equation}
\edit{\item Finally, in the river inlet, $\Gamma_I$:
\begin{equation}
  \label{eq:bc_nondim_last}
  \dyhat{q_c}\Big\rvert_{\Gamma_I} = q_\mathrm{input}'(t) \qquad \text{on $\Gamma_I$}.
\end{equation}}
\end{subequations}
\end{enumerate}

In the above boundary conditions, we have introduced three new dimensionless parameters: $\rho=\frac{r}{K_s}$, $\lambda_s=\frac{L_s}{L_z}$, and $\lambda_c=\frac{L_c}{L_z}$. The river inflow terms in \eqref{eq:river_inflow} \edit{and \eqref{eq:bc_nondim_last}} are converted to non-dimensional form, yielding:
\begin{align}
    \qin &= \mathbf{q_s}\cdot\mathbf{n}\Big\rvert_{\Gamma_R} + \frac{\betazx}{\rho} \int_{\Gamma_R} \mathbf{q_g}\cdot\mathbf{n}\;\d l. \\
    \edit{q_\mathrm{input}'(t)} &\edit{= \left(\frac{\sqrt{S_y}}{n_c}L_c^{5/3}\right)^{-1}q_\mathrm{input}(t)}
\end{align}

Two dimensionless quantities introduced above can be expressed using other quantities:
\begin{equation}
\begin{gathered}
        \lambda_s = \rho \tau_s \quad \text{and} \quad
        \lambda_c = \sqrt{\frac{\gamma\rho\tau_c}{\betazx}}.
\end{gathered}
\end{equation}

Note that we have reduced eleven physical parameters, ($L_x$, $L_y$ $L_z$, $S_x$, $S_y$, $K_s$, $r$, $w$, $n_s$, $n_c$, $\alpha_\mathrm{MvG}$) to nine independent dimensionless parameters ($\betazx$, $\betazy$, $\sigma_x$, $\sigma_y$, $\tau_s$, $\tau_c$, $\gamma$, $\alpha$, $\rho$). This is in agreement with the Buckingham $\pi$ theorem \citep{buckingham1914physically}, which states that the number of dimensionless parameters, $p$, should be equal to $p=n-k$, where $n=11$ is the number of physical variables and $k=2$ is the number of independent physical units (here meters and seconds).

For convenience, in \cref{sec:reduced_models}, we rewrote equations~\eqref{eq:dimless_subsurface} and~\eqref{eq:dimless_overland} in the form:
\begin{subequations}
    \begin{align}
        \frac{\d\theta}{\d h}\bigg\rvert_{h=h_g}\dt{h_g} &=
        \Nop_1(h_g) + \betazy^2 \Nop_2(h_g) + \epsilon \betazy \Nop_3(h_g), \\
        \tau_s\dt{h_s} &= \dxhat{}\left(h_s^{5/3}\right) + R_\mathrm{eff} - I,
    \end{align}
\end{subequations}
where the nonlinear operators $\Nop$ are defined as follows:
\begin{subequations}
    \small
    \label{eq:Nop_definitions}
    \begin{align}
    \label{eq:Nop_definitions_first}
    \Nop_1(h_g) =&
    \begin{aligned}[t]\dzhat{} \left[K_r(h_g) \left(\dzhat{h_g}+1\right)\right]
    + \betazx S_x\dxhat{} \left[K_r(h_g) \left(2\dzhat{h_g}+1\right)\right] \\
    + \betazx^2 \left(1 + S_x^2\right)\dxhat{} \left[K_r(h_g) \dxhat{h_g}\right] - \frac{\d\theta}{\d h}\bigg\rvert_{h=h_g}\dt{h_g},
    \end{aligned} \\
    \Nop_2(h_g) =&  \left(1 + S_y^2\right) \dyhat{} \left[K_r(h_g)\dyhat{h_g}\right], \\
    \Nop_3(h_g) =& 2 \betazx \left(1+S_x^2\right) \dxhat{} \left[K_r(h_g)\dyhat{h_g}\right] + S_x\dyhat{} \left[K_r(h_g) \left(2\dzhat{h_g}+1\right)\right].
    \label{eq:Nop_definitions_last}
    \end{align}
\end{subequations}

\clearpage
\section{List of dimensionless parameters and sizes}
\label{app:list_of_parameters}

\noindent For ease of reference, we include a listing of nondimensional parameters and their typical sizes in \cref{tab:nondim}. 

\vspace*{1.0\baselineskip}
{
\begin{tabular}{lll}
\toprule
    \textsc{parameter} & \textsc{typical size} & \textsc{physical interpretation} \\[1em]
    $S_x$ & $7.5\cdot 10^{-2}$ & slope in the $x$ direction \\[0.2em]
    $S_y$ & $1.4\cdot 10^{-2}$ & slope in the $y$ direction \\[0.2em]
    $\betazx=\frac{L_z}{L_{\hat x}}$ & {\footnotesize\begin{tabular}{@{}l@{}}$5.3\cdot 10^{-3}$ \\ $7.7\cdot 10^{-6}$\end{tabular}} {$\dagger$} & aspect ratio of the cross-section along the hillslope \\[0.2em]
    $\betazy = \frac{L_z}{L_y}$ & {\footnotesize\begin{tabular}{@{}l@{}} $1.1$ \\ $1.6\cdot 10^{-3}$ \end{tabular}} {$\dagger$} & aspect ratio of the cross-section along the channel \\[0.2em]
    $\tau_s = \frac{L_s}{t_0 r}$ & $2.8\cdot 10^{-4}$ & ratio of the overland and groundwater timescales \\[0.2em]
    $\tau_c = \frac{L_c w}{t_0 r L_{\hat x}}$ & $2.9\cdot 10^{-3}$ & ratio of the channel and groundwater timescales \\[0.2em]
    $\gamma = \frac{L_c}{w}$ & $4.0 \cdot 10^{-2}$ & aspect ratio of the stream's cross-section \\[0.2em]
    $\alpha = \alpha_\mathrm{MvG} L_z$ & {\footnotesize\begin{tabular}{@{}l@{}} $2.5\cdot 10^2$ \\ $3.7$ \end{tabular}} {$\dagger$} & dimensionless $\alpha$ parameter from the MvG model \\[0.2em]
\bottomrule
\end{tabular}
\captionsetup{type=table}
\captionof{table}{List of dimensionless parameters. In reference to the mark ($\dagger$), if two values are presented for a single parameter, the top value refers to the \textit{V-shaped catchment} and \textit{deep aquifer} scenarios and the bottom one to the \textit{shallow aquifer} scenario. Otherwise, the parameter value is the same for all scenarios. \label{tab:nondim}}
}

\newpage
\section{Code verification using external benchmarks scenarios}
\label{app:code_verification}

\edit{
\subsection{Test of overland solver}
\noindent Here we test the overland submodel of the numerical solver described in \cref{sec:numerical_methods} using the V-shaped catchment scenario from the intercomparison study by \cite{maxwell2014surface}.}

\edit{In this scenario, the catchment geometry presented in \cref{fig:benchmark_geometries}(a) is used. Only surface flow is allowed, which includes both overland flow along the hillslope and channel flow, each characterised by a different value of Manning's roughness coefficient. In this scenario, a 90-minute rainfall at a uniform intensity of $r=1.8\cdot 10^{-4}~\mathrm{m^3/min}$ is simulated, followed by a 90-minute period of drainage with no rainfall. All numerical values of simulation parameters can be found in \cite{maxwell2014surface}.}

\edit{To test our solver, we compare the hydrograph computed for this scenario with the results from the other coupled surface-subsurface models presented by~\cite{maxwell2014surface}. Since the raw data used to generate the plots are not available, we use an image processing tool in order to reconstruct their data based on the published graphics.}

\begin{figure}
    \centering
    \includegraphics{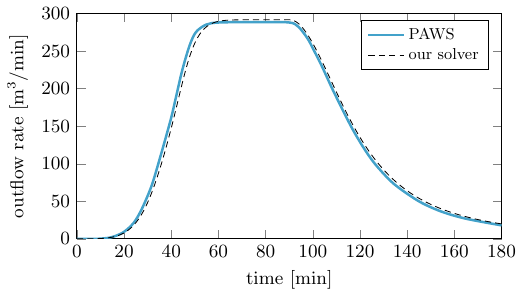}
    \caption{\edit{Comparison of the V-shaped catchment scenario. The solid line represents the hydrograph obtained by \cite{maxwell2014surface} using PAWS, and the dashed lines represent the results obtained by our 3D solver.}}
    \label{fig:V_shape_Maxwell}
\end{figure}

\edit{As shown in \cref{fig:V_shape_Maxwell}, our solver yields almost identical predictions during both the rainfall and drying periods compared to predictions made by PAWS (Process-based Adaptive Watershed Simulator) developed by \cite{shen2010process}. \cite{maxwell2014surface} demonstrated in their intercomparison study that six other tested coupled surface-subsurface software produce similar hydrographs.}

\subsection{Test of 2D solver}

\noindent The coupling of surface and subsurface flow in the numerical model described in \cref{sec:numerical_methods} was tested based on the two-dimensional \textit{saturation excess} and \textit{infiltration excess} scenarios presented in the benchmarking study by \cite{sulis2010comparison}, which were also used in the model intercomparison study by \cite{maxwell2014surface}. 

In both scenarios, we have a catchment constructed from a uniform hillslope (as in \cref{fig:benchmark_geometries}b) made of homogeneous soil, subjected to a constant 200-minute rainfall, followed by a 100-minute period with no precipitation. In the \textit{saturation excess} scenario, the precipitation rate ($3.3\cdot 10^{-4}$ m/min) is lower than the hydraulic conductivity of the soil, allowing the rain to fully infiltrate through the soil, until the soil is fully saturated. In the \textit{infiltration excess} scenario, the precipitation rate is higher than the hydraulic conductivity of the soil. In this case, only a part of the rainwater infiltrates through the ground, while the remaining part forms a so-called \emph{Horton overland flow}.

All the necessary model parameters are presented in the aforemention publications, so no calibration is required. However, information about initial and boundary conditions is missing from the works. We used the same boundary conditions as presented in \cref{sec:Model_formulation}, while for the initial condition, we assumed a constant depth of the groundwater table, with pressure head $h$ decreasing linearly with depth $z$ (this corresponds to no initial vertical flow through the soil).

We compared the results obtained using the finite volume solver described in \cref{sec:numerical_methods} with the results obtained using ParFlow presented by \cite{sulis2010comparison}. As before, we used an image processing tool to extract the data from the graphs presented in these publications.

\begin{figure}
    \centering
    \includegraphics{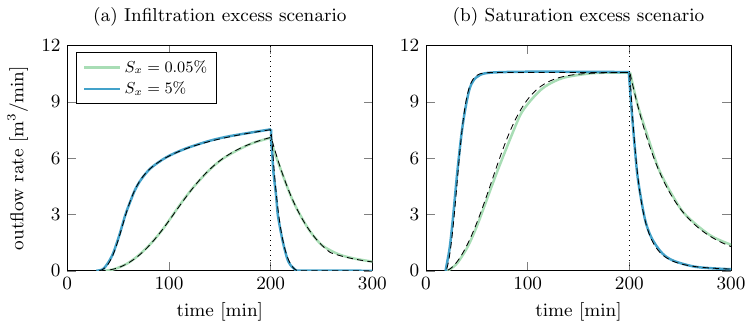}
    \caption{Comparison of (a) infiltration excess ($K_s=6.94\cdot 10^{-5}$ m/min, $\mathrm{wt}=1$ m), and (b) saturation excess scenario ($K_s=6.94\cdot 10^{-4}$ m/min, $\mathrm{wt}=0.5$ m) with two different surface slopes $S_x$. The solid lines represent the hydrograph obtained using ParFlow by \cite{sulis2010comparison}, and the dashed lines represent the results obtained by our 2D solver.}
    \label{fig:Sulis_comparison_plot_1}
\end{figure}

\begin{figure}
    \centering
    \includegraphics{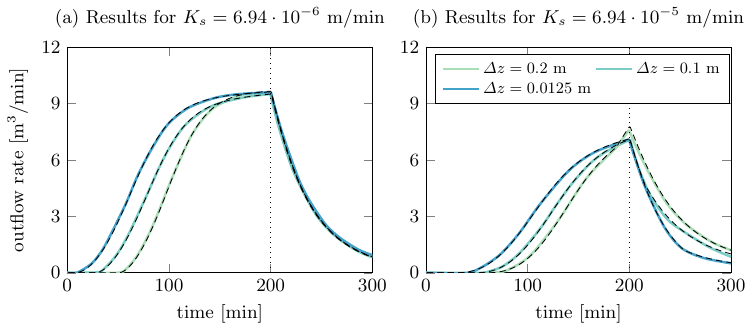}
    \caption{Comparison of infiltration excess scenario with two different vertical mesh resolutions by \cite{sulis2010comparison}. The solid lines represent the hydrograph obtained using ParFlow by \cite{sulis2010comparison}, and the dashed lines represent the results obtained by our 2D solver.}
    \label{fig:Sulis_comparison_plot_2}
\end{figure}

Fig.~\ref{fig:Sulis_comparison_plot_1} demonstrates that the solver very accurately reproduces the results from the original paper in both scenarios for a dense computational mesh ($\Delta z=0.0125$ m). Fig.~\ref{fig:Sulis_comparison_plot_2} additionally shows that the solver also produces almost identical output for lower resolution ($\Delta z=0.1$ m, $\Delta z=0.2$ m), which demonstrates the similarity of our discretisation and numerical artefacts. Since \cite{maxwell2014surface} showed that ParFlow results are consistent with other currently used physical catchment models, we conclude that our solver properly represents all their assumptions within the framework of the considered simple scenario.

\end{document}